\documentclass{pasa}%

\usepackage{times}
\usepackage{graphicx,latexsym, amssymb, amscd, psfrag}
\usepackage{color}
\usepackage{morefloats,rotating,float}
\usepackage{gensymb}
\usepackage{multirow,array}
\usepackage{adjustbox}
\usepackage{gensymb}
\usepackage{mathtools}
\usepackage{framed, caption}
\usepackage{subfigure}
\usepackage{hyperref,breakurl}

\title[Multi-wavelength view of UGC 10420]{Dashing through the cluster: An X-ray to radio view of UGC~10420 undergoing ram-pressure stripping}

\author[Mahajan et al.]{Smriti Mahajan$^1$\thanks{Corresponding author; Email: mahajan.smriti@gmail.com}, Kulinder Pal Singh$^1$, Juhi Tiwari$^1$, Somak Raychaudhury$^{2,3}$
\affil{$^1$Department of Physical Sciences, Indian Institute for Science Education and Research Mohali- IISERM, 
 Knowledge City, Manauli, 140306, Punjab, India }%
\affil{$^2$ Ashoka University, Rajiv Gandhi Education City, Sonepat, Haryana 131029, India }
\affil{$^3$ Inter-University Centre for Astronomy and Astrophysics, Ganeshkhind, Pune, Maharashtra 411007, India }
}%

\def\fuv{{\it FUV }}
\def\ssf{{SFR/$M^*$}}
\def\smass{{$M^*/M_\odot$}}

\def\arcsec{{^{\prime\prime}}}
\def\arcmin{{^{\prime}}}
\def\g{{\it GALEX }}
\def\a{{\it AstroSat }} 
\definecolor{grey}{rgb}{0.5,0.6,0.7}

\definecolor{amber}{rgb}{1.0,0.49,0.0}

\newcommand{\ugc}{{UGC~10420}}

\jid{PASA}
\doi{10.1017/pas.\the\year.xxx}
\jyear{\the\year}

\usepackage{aas_macros}
\usepackage{hyperref} 
\hypersetup{colorlinks,citecolor=blue,linkcolor=blue,urlcolor=blue}

\hypersetup{draft}

\begin{document}

\begin{frontmatter}
\maketitle

\begin{abstract}

 We present multi-wavelength data and analysis, including new \fuv AstroSat/UVIT observations of the spiral galaxy \ugc~($z=0.032$), a member of the cluster Abell~2199. 
 \ugc~is present on the edge of the 
 X-ray emitting region of the cluster at a distance of $\sim 680$ kpc from the centre. The far-ultraviolet ({\it FUV}) data obtained by the {\it AstroSat} mission 
 show intense knots of star formation on the leading edge of the galaxy, accompanied by a tail of the same on the diametrically opposite side. 
 Our analysis shows that  
 the images of the galaxy disk in the optical and mid-infrared are much smaller in size than that in the {\it FUV}.
 While the broadband optical colours of \ugc~are typical of a post-starburst galaxy, the star formation rate (SFR) derived from a
 UV-to-IR spectral energy distribution is at least a factor of nine higher than that expected for a star-forming field galaxy of similar mass at its redshift. A careful removal of the 
 contribution of the diffuse intracluster gas shows that the significant diffuse X-ray emission associated with the inter-stellar medium of \ugc~has a temperature, 
 $T_X = 0.24^{+0.09}_{-0.06}$ keV (0.4--2.0 keV) and luminosity, $L_X = 1.8\pm{0.9}\times 10^{40}$ erg s$^{-1}$, which are typical of the X-ray emission from late-type spiral galaxies. 
 Two symmetrically placed X-ray hot spots are observed on either sides of an X-ray weak nucleus.
 
 Our analysis favours a scenario where the interaction of a galaxy with the hot intra-cluster medium of the cluster, perturbs the gas in the galaxy causing starburst in the 
 leading edge of the disk. On the other hand, the turbulence thus developed may also push some of the gas out of the disk. Interactions between the gas ejected from the galaxy
 and the intracluster medium can then locally trigger star formation in the wake of the galaxy experiencing ram-pressure stripping. Our data however does not rule out the possibility 
 of a flyby encounter with a neighbouring galaxy, although no relevant candidates are observed in the vicinity of \ugc.
 
\end{abstract}
\begin{keywords}
galaxies: evolution; galaxies: fundamental parameters; galaxies: star formation; galaxies: individual: UGC 10420
\end{keywords}
\end{frontmatter}

 \section{Introduction}
 \label{intro}
 
 The evolution of galaxies is often accelerated in the vicinity of galaxy clusters due to the presence of hot intra-cluster medium (ICM), and close proximity to 
 large neighbouring galaxies.
 For several decades astronomers have focused on understanding the various mechanisms responsible for the morphology-density relation \citep{dressler80} and  
 star-formation-density relation \citep[e.g.][]{balogh04} of galaxies. These mechanisms are believed to be the reason for the presence of more red, elliptical and
 passively-evolving galaxies in and near clusters, relative to their counterparts farther away from the cluster centres. 
 
 Gravitational and tidal interactions with comparable-sized neighbours \citep{toomre72, moore96}, emerge as the most important processes leading to the depletion of gas in galaxies 
 residing in intermediate density environments such as the cluster outskirts and large-scale filaments \citep[e.g.][]{porter05, mahajan10, mahajan12}. On the other hand, ram-pressure 
 stripping \citep[RPS;][]{gunn72} plays a significant role in the evolution of galaxies in the dense interiors of clusters and groups. The hydrodynamical interaction of the cold gas in a galaxy 
 with the hot ICM can lead to the evaporation of the interstellar medium \citep[ISM;][]{cowie77}, or complete removal of it from the galaxy via RPS. As a 
 consequence, such a galaxy will eventually cease to form new stars, turn optically red and become elliptical on timescales of $\sim 1$ Gyr. This idea is in agreement with the 
 presence of a large number of HI-deficient galaxies in nearby clusters \citep[e.g.][]{gavazzi06,haines09,vulcani10, boselli14,cybulski14}. 
 
 Hydrodynamical simulations, however suggest that in the initial phase, for a short but significant duration of time RPS can lead to enhancement in the star formation rate of galaxies
 \citep{bekki03,bekki14,steinhauser16,lee20,eagle20}.
 Using multi-wavelength datasets, several incidences of starbursts occurring in RPS galaxies have now been confirmed in nearby clusters 
 \citep[e.g.][]{crowl06,yagi10,gavazzi15,vulcani18,boselli22,lee22}. In a systematic study of RPS candidates at low-redshift, \citet{poggianti16} found that at fixed mass, 
 these galaxies have at least twice the rate of star formation relative to unstripped galaxies.  
  
 The environmental processes such as RPS, which influence the gas content of a galaxy can also impact its nuclear activity. The idea is based on the observations of enhanced star 
 formation among galaxies experiencing RPS. Due to the turbulence and instabilities in the disk, resulting from the interaction of the galaxy with the intra-cluster medium, gas can 
 be funnelled in towards the galaxy's centre. This gas can thereby ignite the central supermassive black hole \citep{tonnesen09}, although observational evidence in 
 favour of these ideas appear debatable at present. In a recent study of nearby galaxies, a high 
 incidence of optical AGN was found to be present among RPS galaxies relative to a mass-matched control sample \citep{peluso22}. However, \citet{cattorini22} suggest that
 the results of \citet{peluso22} may be biased by the fact that they are considering both Seyfert and low-ionization narrow-emission-line region (LINERs) as AGN, 
 and they are evaluating the AGN fraction with respect to the emission-line galaxies only. 
 RPS can also have a significant effect on the surface brightness, temperature and metallicity of the hot interstellar medium of a galaxy \citep{kapferer09}. 
  
 In this paper, we present multi-wavelength data and analysis, including new \fuv observations from AstroSat/UVIT for the RPS galaxy \ugc~which is a member of the cluster Abell 2199 
 ($z=0.031$), and has been observed in many 
 wavebands from X-ray to radio. Some empirical facts about the X-ray cluster Abell 2199, sourced from \citet{rines16} are provided in Table~\ref{a2199}, while physical properties of
 \ugc~compiled from the literature, or derived for this analysis, are listed in Table~\ref{properties}. It is noteworthy that \ugc~has a very small radial offset with the cluster's mean line-of-sight 
 (l.o.s.) velocity, such that $\Delta v \sim 372$ km s$^{-1}$. The structure of the paper is as follows: in the 
 following section we describe various observations and data used in this paper, followed by our analysis 
 in \S~\ref{analysis}. 
 In \S~\ref{results} we discuss the results in the context of the existing literature, and finally conclude the paper with an epilogue in \S~\ref{epilogue}. 
 Throughout this work we use concordance $\Lambda$ cold dark matter cosmological model with $H_0 = 70$ km s$^{-1}$ Mpc$^{-1}$, $\Omega_\Lambda = 0.7$ 
 and $\Omega_m = 0.3$ to calculate distances and magnitudes.

 \section{Observations and data reduction}
 \label{data}
 
 In this section we describe the source and properties of data compiled for \ugc.
 
 \subsection{Ultraviolet data}
 \label{uvdata}
 
 \ugc~was observed with the ultraviolet imaging telescope (UVIT) aboard the \a mission \citep{kumar12, singh14, tandon17}. This field (observation ID: A05\_063T03\_9000003066; 
 PI: Smriti Mahajan; observation date: 30-July-2019) was observed by the UVIT simultaneously in the broadband BaF2 \fuv filter \citep[mean $\lambda = 1541~\AA$; 
 $\Delta\lambda = 380~\AA$;][]{tandon17, tandon20}, and visible wavebands ($3040-5500~\AA$), although the latter data have not been used for scientific analysis here. 
 The raw data from the mission are processed by the Indian Space Science Data Centre (ISSDC), and 
 the Level 1 data from all instruments are provided to the users. A more comprehensive account of processing of data and analysis of this field will be presented in an 
 accompanying paper (Mahajan et al., in preparation), but is briefly mentioned below for completion.

\begin{figure}
 \begin{framed} \centering{
 {\includegraphics[scale=0.4]{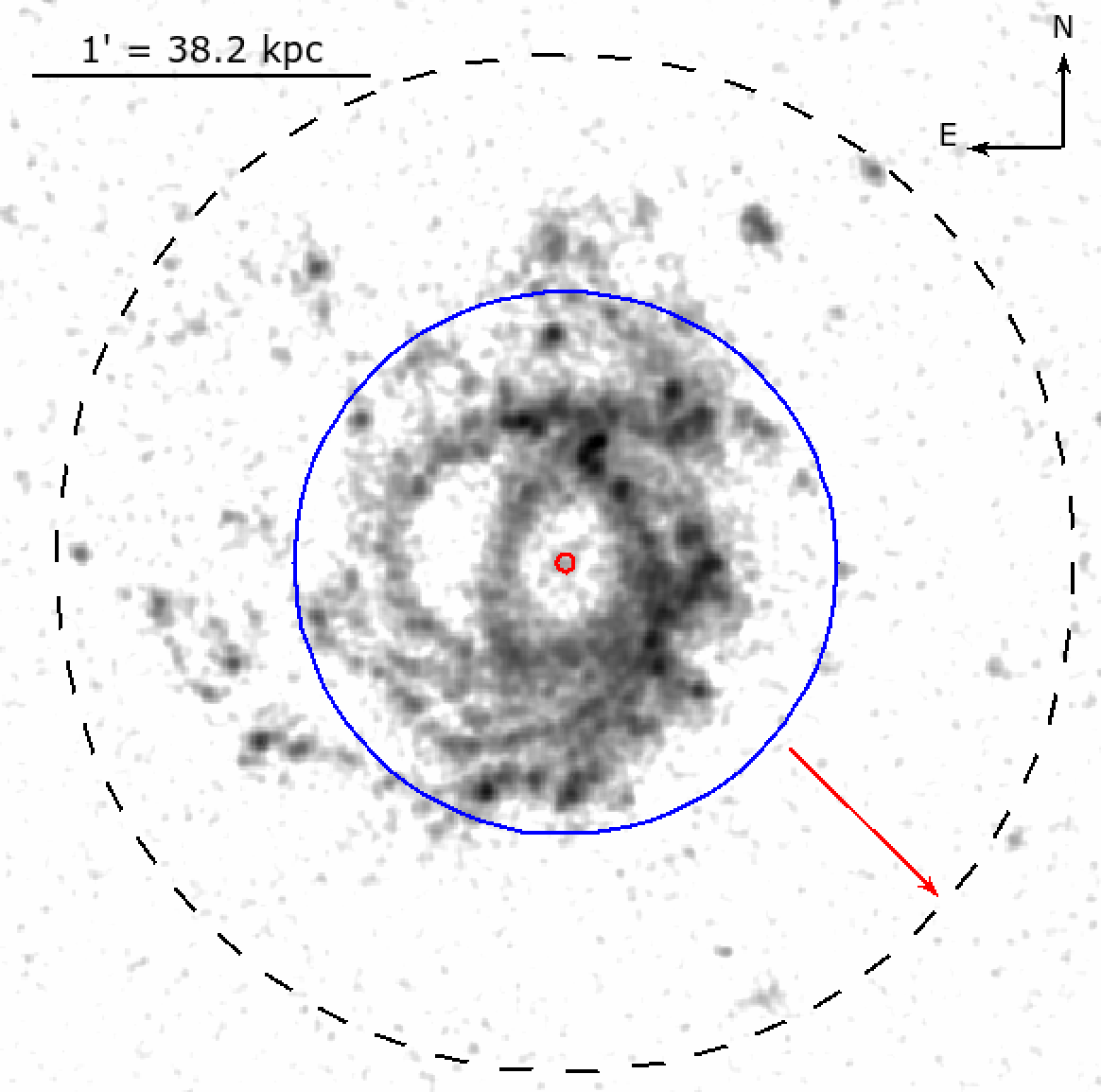}}}
 \end{framed}
 \caption{This is the AstroSat/UVIT image of \ugc. The small {\it red circle} represents the $3\arcsec$ diameter SDSS fibre drawn on the optical centre of the galaxy. 
 The spectrum obtained from this region is shown in Fig.~\ref{spectra}, and represents a broadline emission region. The {\it solid blue} and {\it dashed black} circles having radius of 
 $0.8^{\prime}~\rm{and}~1.5^{\prime}$, respectively are shown for reference to represent the region used for photometric measurements in various filters. The arrow points
 in the direction of the cluster centre.  } 
 \label{uvit}
 \end{figure}

 The UVIT data used in this paper have a total exposure time of 9600.3 seconds and resolution of $<1.2\arcsec$. The L1 data were processed using the CCDLAB 
 software \citep{postma17}, following the procedure described in \citet[][also see \citet{mahajan22} for a step-wise account of creating science images of similar 
 data]{postma21}. The final UVIT \fuv image of \ugc~is shown in Fig.~\ref{uvit}. For comparison, we show two circles of radii $0.8\arcmin$ and $1.5\arcmin$, respectively.
 While the $1.5\arcmin$ region better captures the UV emission of \ugc, the smaller circle is a better representation of the size of the galaxy at longer wavebands,
 as shown below. 
 
 \begin{figure*}
 \centering{
 {\includegraphics[scale=0.455]{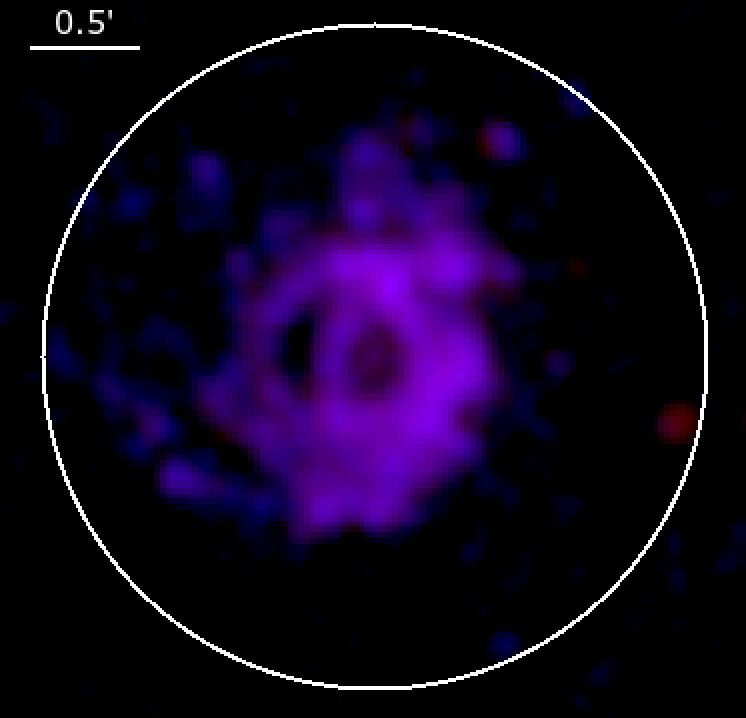}}}
 \centering{
 {\includegraphics[scale=0.36]{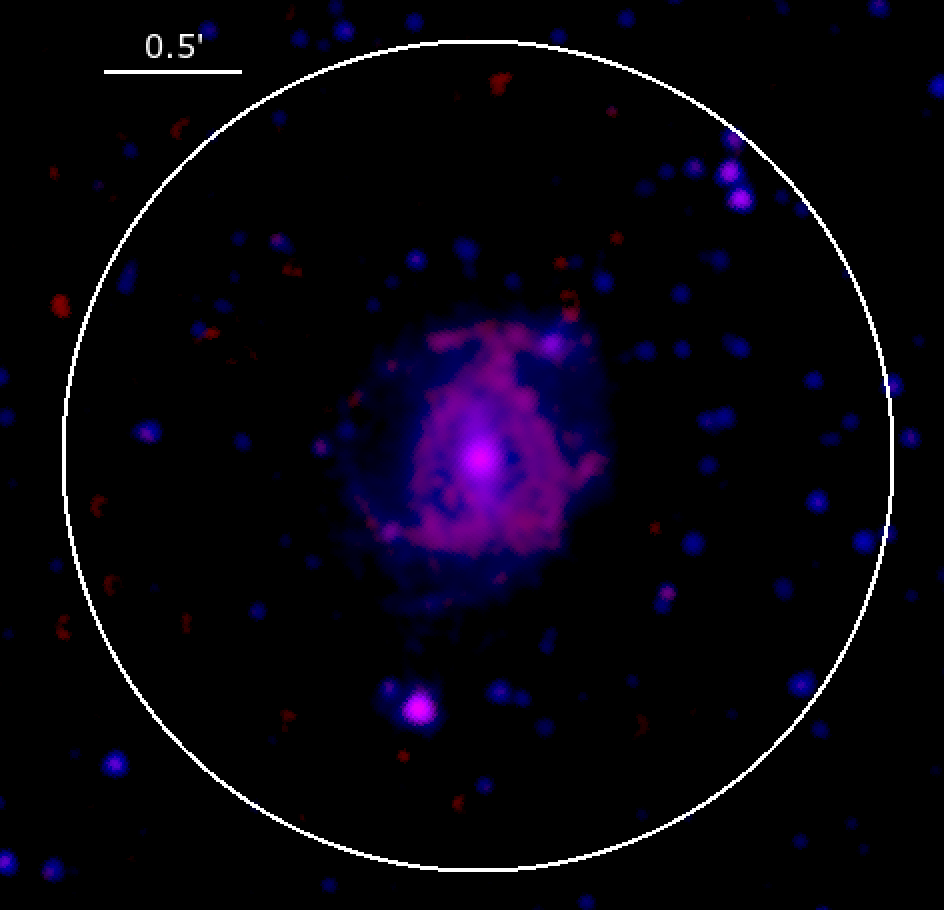}}}
 \caption{ {\it (left:)} The \g~image of \ugc~where {\it blue} and {\it red} represent the {\it FUV} and {\it NUV} broadband images, respectively. {\it (right:)} This  
 composite image of \ugc~shows the NIR emission in the IRAC 3.6 $\mu$m {\it (blue)} and 5.8 $\mu$m {\it (red)} bands. A circle of $1.5\arcmin$ radius is shown 
 in both the images for comparison. It is evident that the image of the galaxy is much smaller in the infrared bands.     } 
 \label{irac-galex}
 \end{figure*}

 \ugc~was also observed by the Galaxy and Mass Evolution Explorer ({\it GALEX}) in the near and far ultraviolet broadbands as shown in 
 Fig.~\ref{irac-galex} (left). It is noteworthy that the shallow exposure time and poor resolution renders this image unimpressive relative to the UVIT counterpart shown 
 in Fig.~\ref{uvit}. These data were taken as part of {\it GALEX}'s medium imaging survey (MIS), and have a total exposure time of 1649.1 seconds in both the bands. 
 
 In order to measure the flux from \ugc, an aperture had to be chosen manually because the commonly used software split the galaxy into multiple components due to 
 its large size, and discontinuous emission in almost all filters. 
 The aperture photometry was performed using the SAOImageDS9 software package \citep{ds9}, by choosing an aperture size of $1.5\arcmin$ centred at the optical 
 nucleus ($\alpha$: 16h29m51.04s; $\delta$: +39d45m59.42s) for all three UV images. First, we estimated the background counts by averaging over multiple source-free circular regions of 
 radius $\geq 1.5\arcmin$, and subtracted them from the counts obtained for the region centred on \ugc. The counts were then converted to fluxes using the standard 
 conversion factors given in \citet{tandon20} for the UVIT image, and \url{https://asd.gsfc.nasa.gov/archive/galex/FAQ/counts_background.html} for the \g~images, 
 respectively. The final flux values, measurement uncertainty in flux, and the effective wavelength for all wavebands are listed in Table~\ref{tab:sed}.

  \begin{figure*}
    \centering
    \begin{subfigure}
    \centering
    \includegraphics[width=0.45\linewidth]{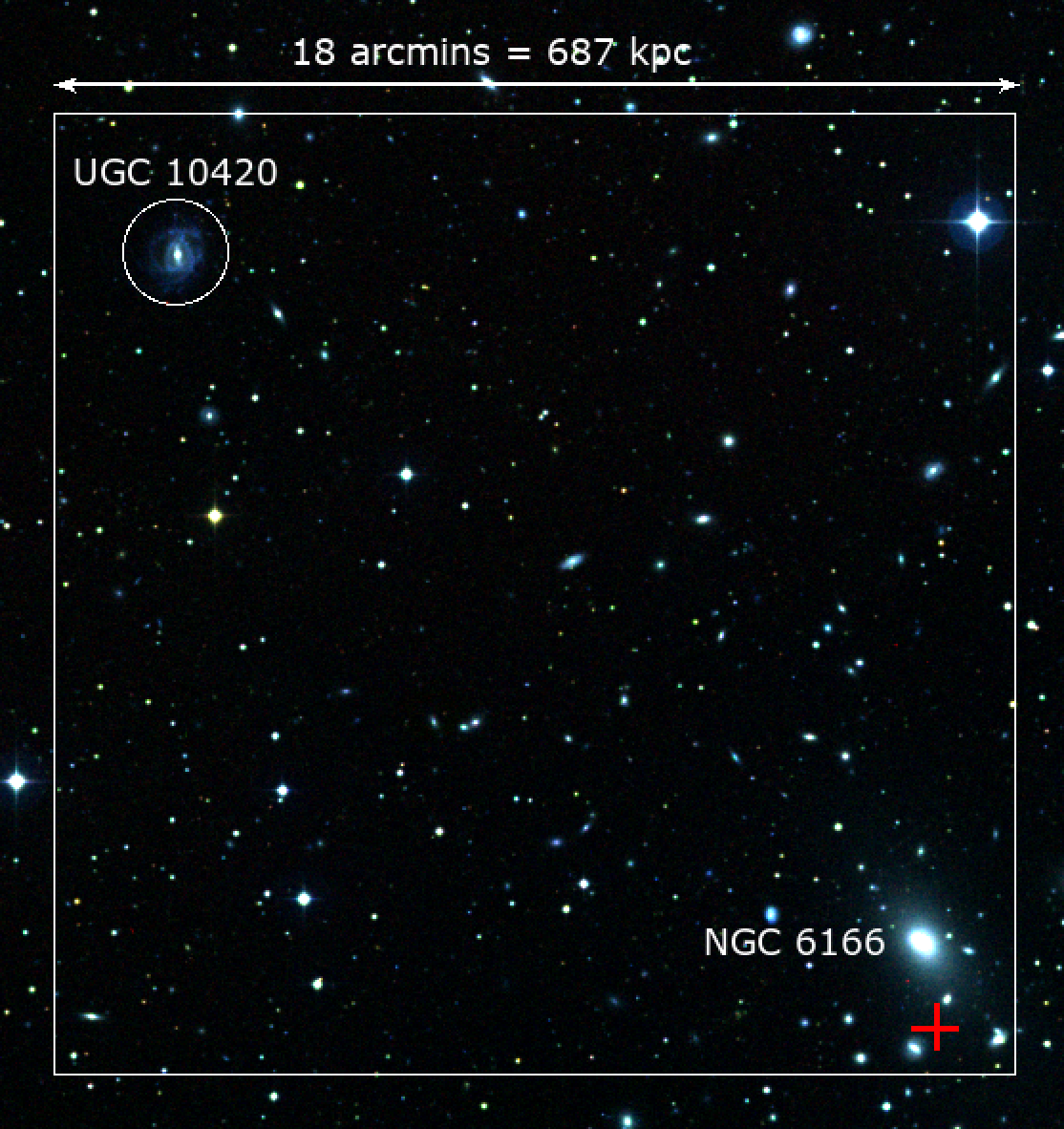}
    \end{subfigure}
    \begin{subfigure}
    \centering
    \includegraphics[width=0.5\linewidth]{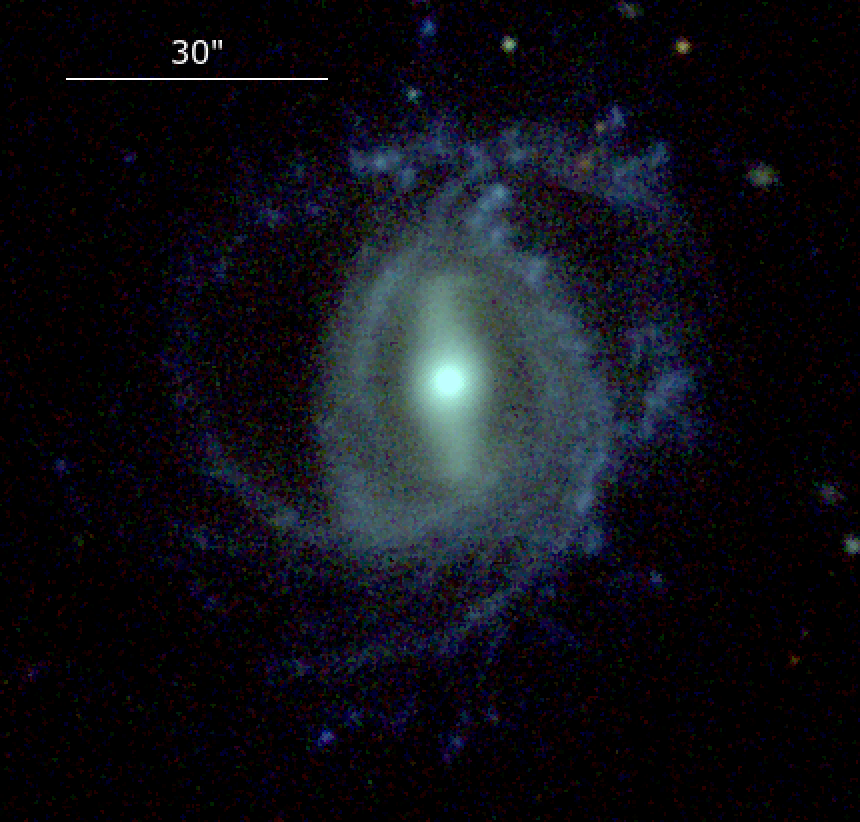}
    \end{subfigure}  
    \caption{{\it (left:)} The combined $B, R$ and $I$ band digitized sky survey (DSS) image of the sky around \ugc. The central elliptical galaxy of Abell 2199, NGC 6166 
    is marked along with the X-ray centre of the cluster \citep[{\it red cross};][]{mahdavi01} at the bottom right to show the relative position of \ugc~in the cluster. 
    {\it (right:)} This is an optical $g,r,i$ image of \ugc~created using the SDSS photometric data. North is up and east is on the left for both images.}
    \label{gri}
 \end{figure*}

 \begin{table*}
 \caption{Integrated flux measurements for \ugc~in different wavebands, in increasing order of wavelength from top to bottom. }
 \begin{center}
 \begin{tabular}{ lrrrl }     
 \hline
 Filter & Effective $\lambda$  & Flux  & Uncertainty in flux  & Source \\ 
         &  (\AA\ )                    & (mJy)      &    (mJy)     &         \\
 \hline
\g $FUV$	 	&	0.152	&	1.486		&	4.00E-01		&   measured from image	\\
UVIT BaF2	&	0.154	&	2.030		&	8.04E-03		&   measured from image	\\
\g $NUV$		&	0.232	&	2.315		&	2.90E-01		&  measured from image	\\
 SDSS $u$	&	0.356	&	2.573		&	9.66E-02		&  measured from image	\\
$U$			&	0.357	&	3.060		&	2.64E-00		&	from \citet{gildepaz07}	\\
$B$			&	0.446	&	6.877		&	5.99E-00		&	from \citet{gildepaz07}	\\
SDSS $g$		&	0.477	&	6.920		&	1.58E-01		&  measured from image	\\
$V$			&	0.550	&	11.197		&	9.66E-00		&	from \citet{gildepaz07}	\\
SDSS $r$		&	0.623	&	11.410		&	2.03E-01		&  measured from image	\\
SDSS $i$		&	0.763	&	15.438		&	2.37E-01		&  measured from image	\\
SDSS $z$		&	0.913	&	18.375		&	2.58E-01		&  measured from image	\\
$J$			&	1.200	&	18.370		&	17.87E-00		&	from \citet{gildepaz07}	\\
$H$			&	1.620	&	19.653		&	18.77E-00		&	from \citet{gildepaz07}	\\
$K$			&	2.200	&	12.767		&	11.86E-00		&	from \citet{gildepaz07}	\\
WISE $W1$	&	3.368	&	12.922		&	2.84E-02		&  measured from image	\\
IRAC $I1$		&	3.560	&	14.254		&	3.76E-01		&  measured from image	\\
WISE $W2$	&	4.618	&	7.653		&	2.14E-02		&  measured from image	\\
IRAC $I3$		&	5.760	&	14.457		&	9.95E-01		&  measured from image	\\
WISE $W3$	&	12.082	&	13.993		&	0.17E-00  		&  measured from image	\\
WISE $W4$	&	22.194	&	16.027		&	0.75E-00		&  measured from image	\\
Radio 144 MHz	 &	2.087E+10	&	74.670	&	0.17E-00		&  measured from image	\\ 
\hline
\end{tabular}
 \end{center}
 \label{tab:sed}  
 \end{table*}

 \subsection{X-ray Data}
 \label{xdata}
 
\ugc~was in the field of view of an observation with the \textit{XMM-Newton} EPIC\footnote{The European Photon Imaging Camera onboard the XMM-Newton observatory.}. This observation (Obs. ID: 0784521201, PI: Takayuki Tamura, Obs. time: 40 ks) carried out on 2017 March 17
was centred at $\alpha$: 16h30m32s and $\delta$: +39d44m52s and focused on the intermediate regions ($< 3/4$ virial radius) of Abell 2199. All the three detectors aboard \textit{XMM-Newton}, viz., EPIC PN, MOS1 amd MOS2 detectors were operated in the Full Frame mode and medium optical blocking filter was used. The dataset for this observation was retrieved from the HEASARC\footnote{High Energy Astrophysics Science Archive Research Center; \url{https://heasarc.gsfc.nasa.gov/}} archive, then processed and analysed using the \textit{XMM-Newton} Science Analysis System (\texttt{SAS}) version 20.0.0 and the Extended Source Analysis Software (\texttt{XMM-ESAS}) package integrated into \texttt{SAS}. The latest calibration files were employed to produce the PN (MOS) event files using the \texttt{SAS} tasks \textit{epchain} (\textit{emchain}). 

 The X-ray data were first screened to remove the particle backgrounds as described below:

a) The event files were filtered for soft proton (SP) background flares using the tasks \textit{pn-filter} and \textit{mos-filter}. The SP flare filtering resulted in useful exposure times of 28.92, 35.46, and 38.20 ks for the PN, MOS1, and MOS2 detectors, respectively. We examined the output of the task \textit{mos-filter} and verified that none of the MOS detectors operated in an anomalous state during the observation. \\
b) The events arising due to quiescent particle background (QPB) produced by the interaction of high energy particles ($> 100$ MeV) with the X-ray detectors and the surrounding structure was then removed. QPB shows only small intensity variations over the time scale of a typical X-ray observation, and therefore not detected as flares. The standard norm to subtract out the QPB contribution from the data is to either use blanksky observations or model the QPB. The \texttt{XMM-ESAS} package models the QPB using a combination of a database of unexposed-region data and filter-wheel-closed data based on the methods of \cite{kuntz2008}, and we have used this method to model and subtract the QPB instead of using blanksky files.\\

The filtered event files thus produced were used for making the images and spectra, and analysed as described below.

\begin{figure*}
\begin{centering}
\includegraphics[scale=0.24]{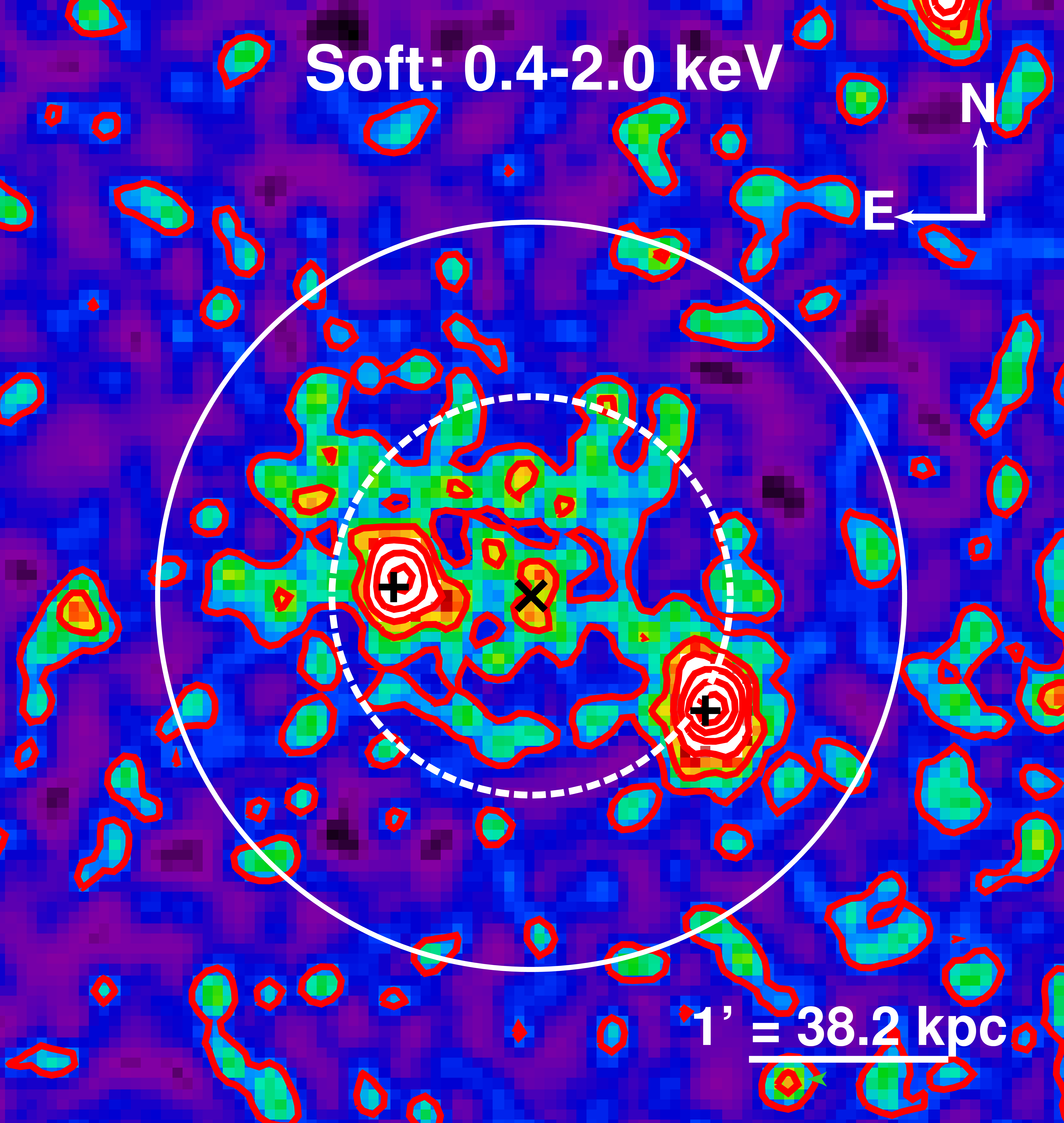}
\includegraphics[scale=0.24]{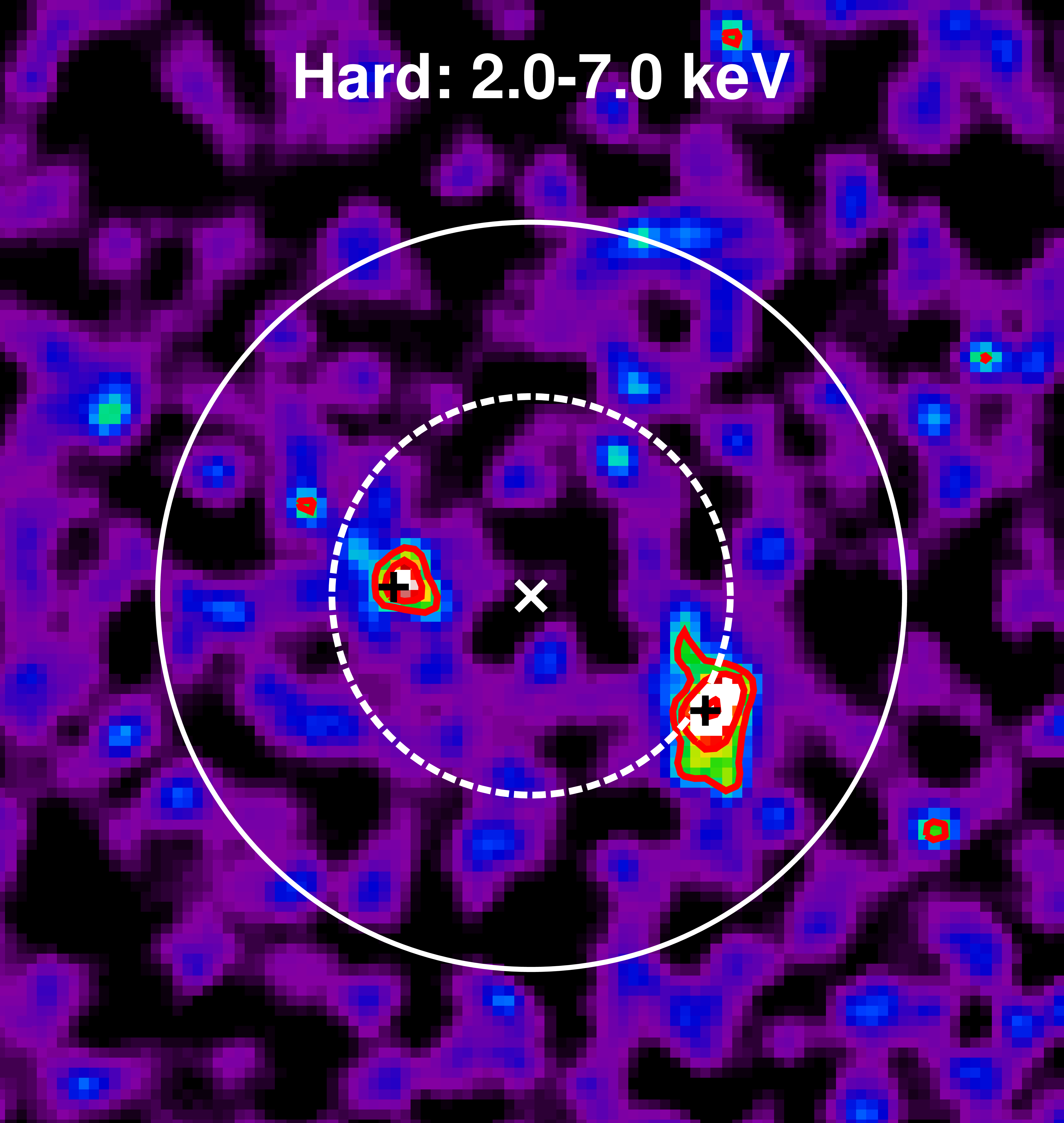}
\includegraphics[scale=0.24]{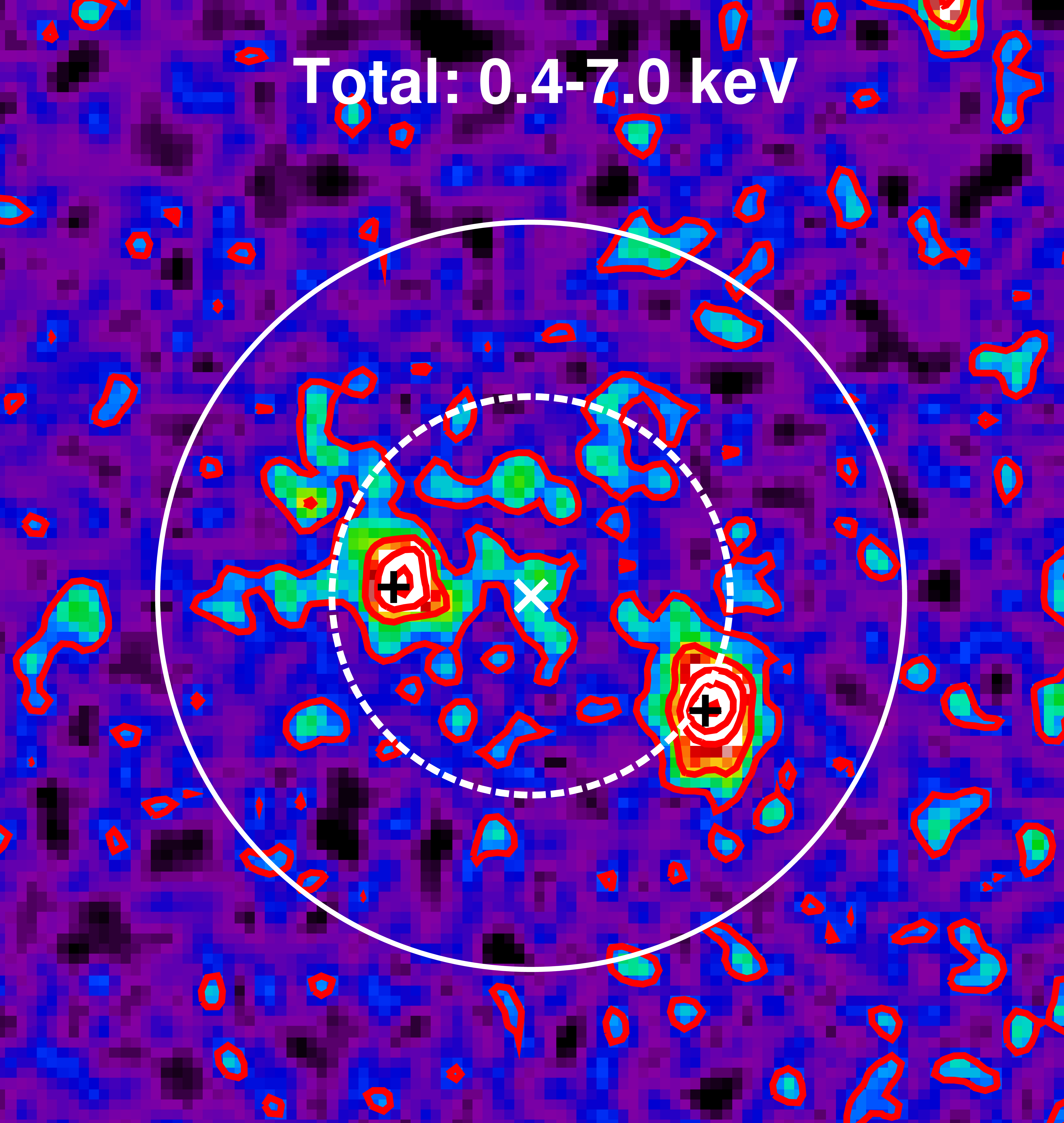}
\caption{\textit{left to right}: Particle-background-subtracted, exposure-corrected, combined PN, MOS1, and MOS2 \textit{XMM-Newton} surface brightness image of UGC 10420 in the \textit{soft} (0.4--2.0 keV), \textit{hard} (2.0--7.0 keV), and \textit{total} (0.4--7.0 keV) X-ray energy bands. The surface brightness contours are as follows -- \textit{soft band}: 6 levels (9.9, 20.9, 36.1, 61.1, 105.0, and 184.3 cts s$^{-1}$ deg$^{-2}$) between 3$\sigma$ and $88\sigma$ above a background level of 3.7 cts s$^{-1}$ deg$^{-2}$; \textit{hard band}: 3 levels (17.1, 30.8, and 81.3 cts s$^{-1}$ deg$^{-2}$) between 3$\sigma$ and 19$\sigma$ above a background level of 5.1 cts s$^{-1}$ deg$^{-2}$; \textit{total band}: 5 levels (16.5, 43.9, 87.4, 172.8, and 350.2 cts s$^{-1}$ deg$^{-2}$) between $3\sigma$ and $98\sigma$ above a background level of 6 cts s$^{-1}$ deg$^{-2}$. In each panel, the $\times$ symbol marks the optical centre of UGC 10420, and the $+$ symbols mark the centres of the two X-ray point sources detected in the galaxy. The \textit{solid white} and \textit{dashed white} circles represent the $1.5\arcmin$ and $0.8\arcmin$ radius apertures, respectively, centred at the optical centre.}
\label{fig:ugc10420xray}
\end{centering}
\end{figure*}

The \texttt{XMM-ESAS} tasks \textit{pn-spectra} (\textit{mos-spectra}) followed by \textit{pn$\_$back} (\textit{mos$\_$back}) were used to generate the PN (MOS1/MOS2) images of UGC 10420 in three different energy bands (one at a time), viz. soft: $400-2000$ eV, hard: $2000-7000$ eV, and total: $400-7000$ eV, by setting the \textit{elow} and \textit{ehigh} energy parameters to the requisite values (unit: eV). These tasks also generated the QPB images and exposure maps for each EPIC detector. The individual PN, MOS1, and MOS2 counts images, exposure maps, and model QPB images were then combined into a single counts, exposure, and QPB image using the task \textit{comb}. Finally, a combined PN, MOS1, and MOS2 particle-background-subtracted, exposure-corrected, and adaptively smoothed surface brightness image of UGC 10420 was generated using the task \textit{adapt}. A smoothing scale of 10 counts and a pixel size of 
$2.5\arcsec$ were used.  X-ray images of UGC 10420 in the three energy bands are shown in \autoref{fig:ugc10420xray}. The nuclear region around the optical centre appears to be very faint and soft in X-rays, with two bright spots, neither of which is coincident with the optical centre, and surrounded by some diffuse emission. Emission from the central region cannot be seen in the hard X-ray band, however. The two bright sources far away from the nuclear region on the east and west are confirmed as point sources using the ESAS task \textit{cheese}, which runs point source detection by calling the SAS task \textit{edetect$\_$chain}.  We used a detection box of $5\times 5$ pixels for this purpose. Further analysis of these X-ray sources is presented in \S3.3.

 \begin{table}
 \caption{Empirical facts about Abell 2199 \citep{rines16}. }
 \begin{center}
 \begin{tabular}{ ll }     
 \hline
 Parameter & Value \\ 
  \hline
 Right Ascension (J2000)	& 16 28 37.90 	\\
 Declination (J2000)	& +39 32 55.32		 \\
 Redshift ($z$)	&	0.0309  	 	\\
 Velocity dispersion ($\sigma$) & $676^{+37}_{-32}$ km s$^{-1}$ \\
 $M_{200}$  & $(2.39 \pm 0.77) \times 10^{14}$ M$_\odot$ \\
 \hline
\end{tabular}
 \end{center}
 \label{a2199}  
 \end{table}

  \begin{table*}
 \caption{Observed properties of \ugc, a member galaxy of cluster Abell 2199. }
 \begin{center}
 \begin{tabular}{ lll }     
 \hline
 Parameter & Value & Reference \\ 
 \hline
 Right Ascension (J2000)	&	16 29 51.04	&	SDSS database \\
 Declination (J2000)	&	+39 45 59.42	&	SDSS database \\
 Redshift ($z$)	&	0.03185	&	SDSS database \\
 Luminosity distance ($D_L$) 	& 	139.7 $h^{-1}$ Mpc 	& -	\\
 Scale (kpc/$^{\prime\prime}$) 	& 	0.636 	& - 	\\
 Morphological type	&	SB(c)	&	 \citet{buta}	\\
 Position angle (PA) of stellar bar	&  $89.9^{\circ}$	& \citet{fraser20}	\\
 Length of stellar bar   & 3.9  kpc	& \citet{fraser20}	\\
Petrosian 50\% radius in $r$-band, $R_{50,r}$	 & $18.2\arcsec$	& \citet{fraser20}	\\
 Milkyway extinction, $E(B-V)$	&  0.0085	&  \citet{schlafly11} \\
 \hline
 \end{tabular}
 \end{center}
 \label{properties}  
 \end{table*}

 \subsection{Optical data} 
 \label{optical}
 
 The optical photometric and spectroscopic data used in this paper are sourced from the Sloan Digital Sky Survey \citep[SDSS, data release 14;][]{sdss14}, unless 
 mentioned otherwise. The SDSS images have an exposure time of 53.9 seconds in each of the five filters $u,g,r,i,z$. Photometry was performed in all five wavebands
 using an aperture size of $0.8\arcmin$, following the procedure described in Sec.~\ref{uvdata}. The measured counts were then converted to fluxes using the standard conversion 
 factor from maggies to Jy, and are listed in Table~\ref{tab:sed}. 
  
 In Fig.~\ref{gri} (left), we show the position of the galaxy relative to the centre of Abell 2199, on the digitized sky survey (DSS) image of the field. \ugc~is $\sim 680$ kpc away
 from the cD galaxy NGC 6166 of the cluster. A composite $g,r,i$ image of \ugc, shown in Fig.~\ref{gri} (right), clearly shows the asymmetry in the spiral 
 arms of \ugc, and the presence of a bar at the centre, in agreement with the morphological classification of type SB(c) \citep{buta}. 
 Even though the outer regions of \ugc~are optically blue, majority of the optical emission from the disk is dominated by red stellar populations, as 
 quantified by a colour value of $(g-r) \sim 0.8$ mag, which is typical for passively-evolving late-type galaxies \citep[e.g.][]{mahajan09, song17, mahajan20}.    
 
\begin{figure}
  \centering{
 {\includegraphics[scale=0.3]{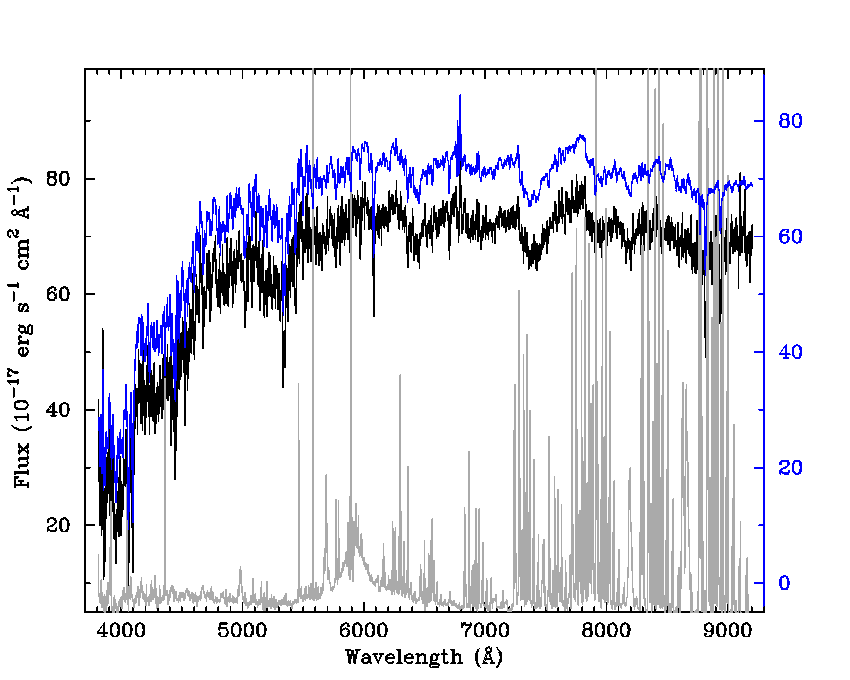}}}
 \caption{The spectrum {\it (black line)} from the nuclear region of \ugc~(also see Fig.~\ref{uvit}), based on which it is classified as a LINER galaxy. 
  The {\it blue} and {\it grey} lines represent the best-fit model to the spectrum, and emission from the sky, respectively. The model spectrum is offset along the ordinate for clarity, 
  and the corresponding shifted axis is shown in {\it blue} on the right. All the data shown in this image are sourced from the SDSS database. } 
  \label{spectra}
  \end{figure}

 The optical spectrum of \ugc~obtained from the $3\arcsec$ diameter SDSS fibre is shown in Fig.~\ref{spectra} (also see Fig.~\ref{uvit} for a comparison between the 
 size of the fibre relative to the galaxy). On the basis of emission line fluxes obtained from this spectrum,  \ugc~is classified as a LINER galaxy, using the BPT \citep{bpt} 
 emission-line diagnostic diagram \citep{toba14}.

 \subsection{Infrared data} 
 
 The infrared (IR) observations for \ugc~were obtained from the Infrared Science Archive (IRSA) of the NASA Infrared Processing and Analysis Center 
 (IPAC)\footnote{https://irsa.ipac.caltech.edu/about.html}. For this work we used the images from the Wide-field Infrared Survey Explorer \citep[WISE;][]{wright10}, 
 as well as the Infrared Array Camera (IRAC) aboard the {\it Spitzer} mission. 
 
\begin{figure}
 \centering{
 {\includegraphics[scale=0.5]{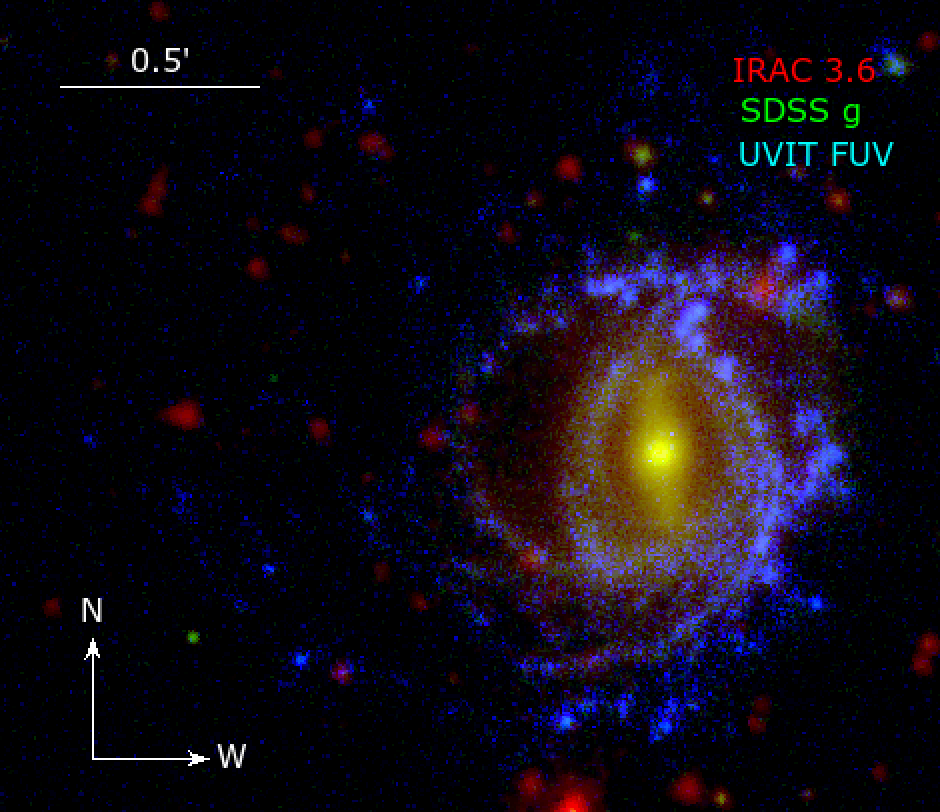}}}
 \caption{A composite image of \ugc~created using the IRAC 3.6 $\mu$m {\it (red)}, SDSS $g$-band {\it (green)} and UVIT {\it FUV} {\it (blue)} data, respectively. Despite 
 different pixel sizes and depth of the three datasets, this image shows the distorted spiral arms of \ugc which are likely perturbed by the interaction with the intra-cluster medium. 
 It is also evident that most of the star-formation is now occurring in the outer regions of the galaxy. } 
 \label{rgb}
 \end{figure}

 WISE observed the sky in the near- and mid-infrared wavebands between 3.6 to 22 $\mu$m. 
 We performed photometry on all the four band {\it ALLWISE} images, and channels 1 and 3 IRAC images (3.6 and 5.8 $\mu$m, respectively) in the same manner as 
 described in sec.~\ref{uvdata}, but using an aperture of $0.8\arcmin$. 
 Our choice of aperture size is dictated by the fact that the \ugc's image is much smaller in these bands, and, the aperture of $1.5\arcmin$ used for the UV data
 would include many galactic and extragalactic sources not visible in other bands, thereby contaminating the flux measurement.
 The estimated fluxes are listed in Table~\ref{tab:sed}. A glimpse of the IR emission relative to the light in the optical and {\it FUV}
 filters can be seen in Fig.~\ref{irac-galex} (right) and Fig.~\ref{rgb}. Figs. 1--3 and 5 show that the optical-IR emission is concentrated in the central region of the galaxy encompassing the 
 nucleus and the stellar bar, where as the {\it FUV} emission dominates the outer regions of the galaxy. 
  
 \subsection{Radio data} 
 
 The radio image of \ugc~is obtained from the data release 2 of the ongoing LOw-Frequency ARray (LOFAR) Two-metre Sky Survey  
 \citep[LoTSS; 120--168 MHz;][]{shimwell22}.
 The LoTSS images were derived from the LOFAR high band antenna data, which were corrected for the direction-independent instrumental properties as well as 
 direction-dependent ionospheric distortions with fully automated, data processing. At $\sim 6\arcsec$ resolution, the full bandwidth Stokes I continuum maps of these data at
 a central frequency of 144 MHz, have a median root mean square (rms) sensitivity of $83~\mu$Jy/beam, a flux density scale accuracy of approximately 10\%, and astrometric 
 accuracy of $0.2\arcsec$. 
 
\begin{figure}
 \centering{
 {\includegraphics[scale=0.4]{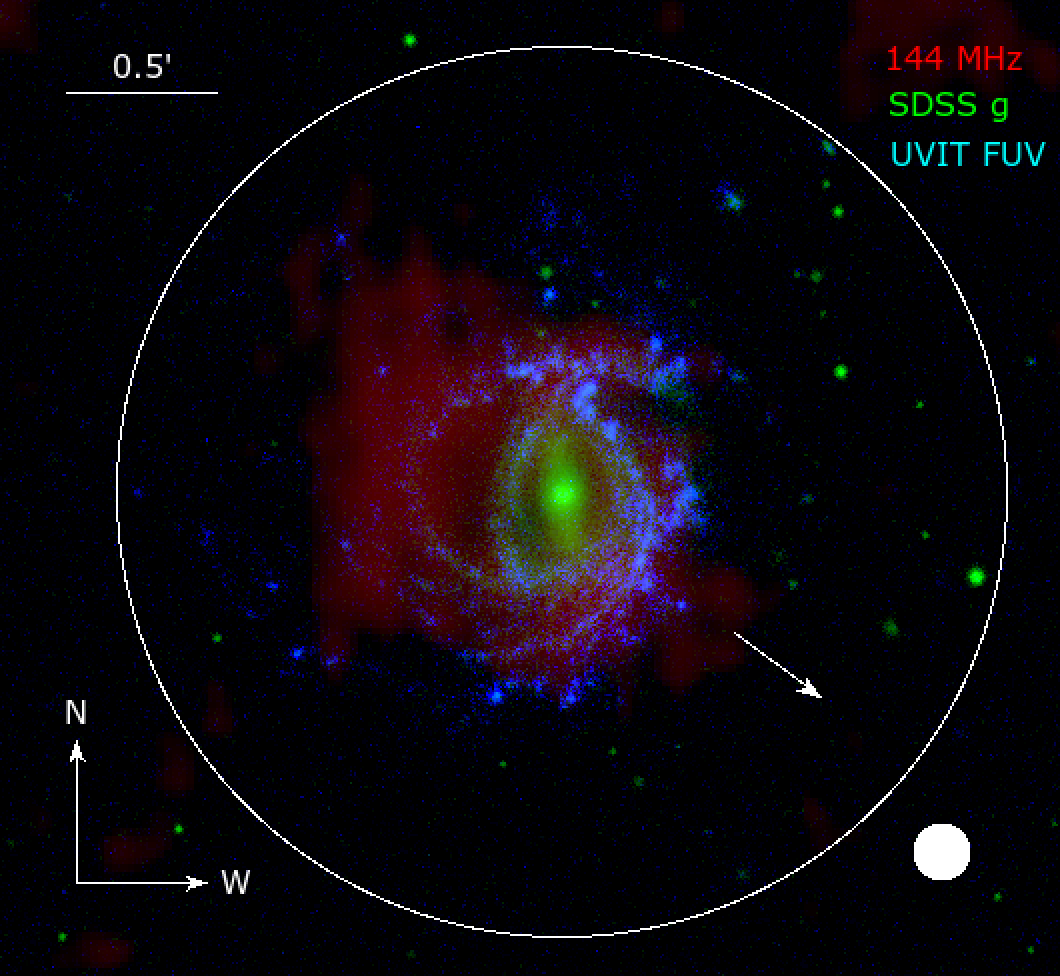}}}
 \caption{This composite image of the \ugc~shows the low-frequency LOFAR radio emission {\it (red)}, the optical $g$-band {\it (green)} and the UVIT {\it FUV} emission {\it (blue)},
 respectively. The synthesised radio beam is represented by the $6\arcsec$ radius circle at the bottom right. 
 It is noteworthy that an excess of radio emission is detected in the northwestern part of the galaxy, opposite to the direction of the cluster centre. 
 The arrow points in the direction of the cluster centre.   } 
 \label{rad-g-uvit}
 \end{figure}

 The radio flux density mentioned in Table~\ref{tab:sed} was obtained using {\sc aips}. In order to determine the flux density, the counts in a polygon region centred 
 at \ugc~were determined and divided by the area covered by the beam. The uncertainty in the flux was calculated as the product of the square root of the number 
 of beams within the polygon region, and the rms uncertainty ($=0.017$ mJy) in a similar sized source-free region. 
 The flux density scale uncertainty for the LoTSS data is found to be $\sim 6\%$ \citep{shimwell22}. 
 The 144 MHz emission encompasses the optical disk of the galaxy as shown in an optical-UV-radio composite image of \ugc~in Fig.~\ref{rad-g-uvit}. The radio emission appears to be 
 orientated along the direction of motion of the galaxy with a gap in the middle, in line with the {\it FUV} image of the galaxy (Fig.~\ref{uvit}).
 
 \section{Analysis and Results}
 \label{analysis}
 
 \subsection{Spectral energy distribution}

 \begin{figure}
 \centering{
 {\includegraphics[scale=0.35]{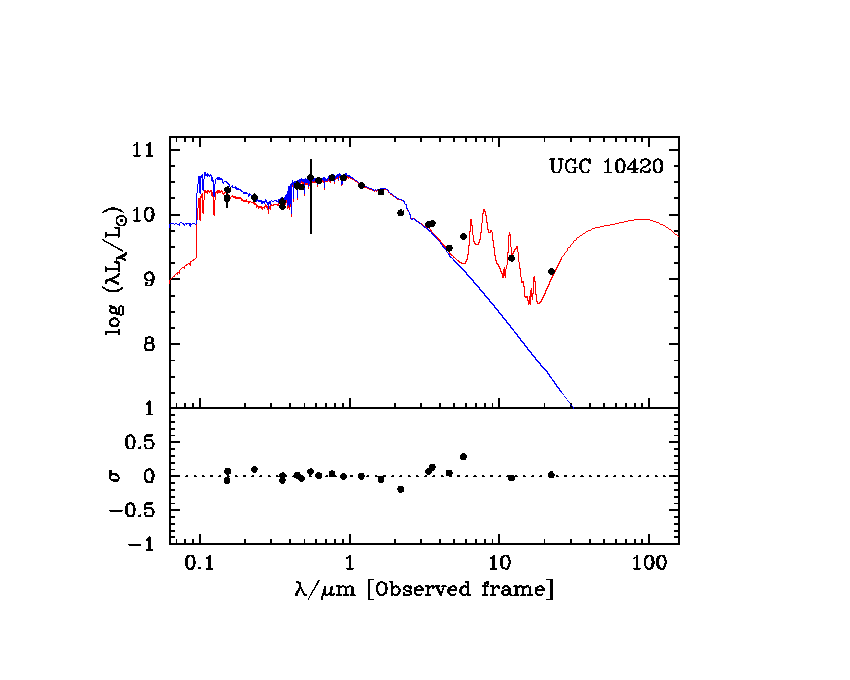}}}
 \caption{The integrated UV to near-infrared spectral energy distribution (SED) of the galaxy \ugc. In the {\it top panel}, photometric data from Table~\ref{sed}
 are shown as {\it black points}. These data are used to fit an SED using the code {\sc magphys} \citep{dacunha08}. The attenuated {\it (red line)}, and unattenuated {\it (blue line)}
 best-fit SEDs are also shown. The residuals with respect to the best-fit are shown in the {\it bottom panel}. Also see Table~\ref{magphys} for the physical parameters derived 
 from this SED.    } 
 \label{sed}
 \end{figure}

 The arsenal of data mentioned in Section~\ref{data} can be used to derive useful information about the stars, gas and dust of \ugc~by means of its spectral energy distribution 
 (SED). In this work we use the self-contained model package
 called Multi-wavelength Analysis of Galaxy Physical Properties \citep[{\sc magphys};][]{dacunha08} to model the SED of \ugc. {\sc magphys} takes the photometric data from UV to IR 
 wavebands as input, to model the attenuated and unattenuated SED of a galaxy. 
 It makes use of the stellar evolution models by \citet{bruzual03}, assuming a \citet{chabrier03} initial mass function and a \citet{charlot00} attenuation law.
 For \ugc~we input the data listed in Table~\ref{tab:sed} (sans radio flux). 
 
 The output is shown in Fig.~\ref{sed}, and some of the derived physical parameters: star formation rate (SFR)\footnote{Since this SFR is based on \citet{chabrier03} IMF, we have divided  
 by 0.63 \citep{madau14} to convert to the Salpeter IMF used throughout this work.}, stellar mass ($M^*$), dust mass (M$_{Dust}$) and specific star 
 formation rate (SFR/$M^*$; sSFR henceforth) are listed in Table~\ref{magphys}, respectively.
The physical properties of \ugc~are typical of optically-selected red spiral galaxies explored in the literature \citep[e.g.][]{mahajan20}. 

 \begin{table}
 \caption{Some physical parameters estimated from the SED of \ugc~fitted using the {\sc magphys} code \citep{dacunha08}. }
 \begin{center}
 \begin{tabular}{ ll }     
 \hline
 Parameter & Value \\ 
  \hline
 SFR, $M_{\odot}$ yr$^{-1}$ & $9.75$  \\
 Stellar mass, \smass & $1.8 \times 10^{10}$  \\
  Dust mass, $M_{Dust}/M_\odot$ &  $6.8 \times 10^6$ \\
 \ssf  & $3.4 \times 10^{-10}$  \\
  \hline
\end{tabular}
 \end{center}
 \label{magphys}  
 \end{table}

 \subsection{Star forming regions}
 \label{s:regions}
 
 \begin{figure*}
 \centering

  \frame{\subfigure[]{\includegraphics[scale=0.3]{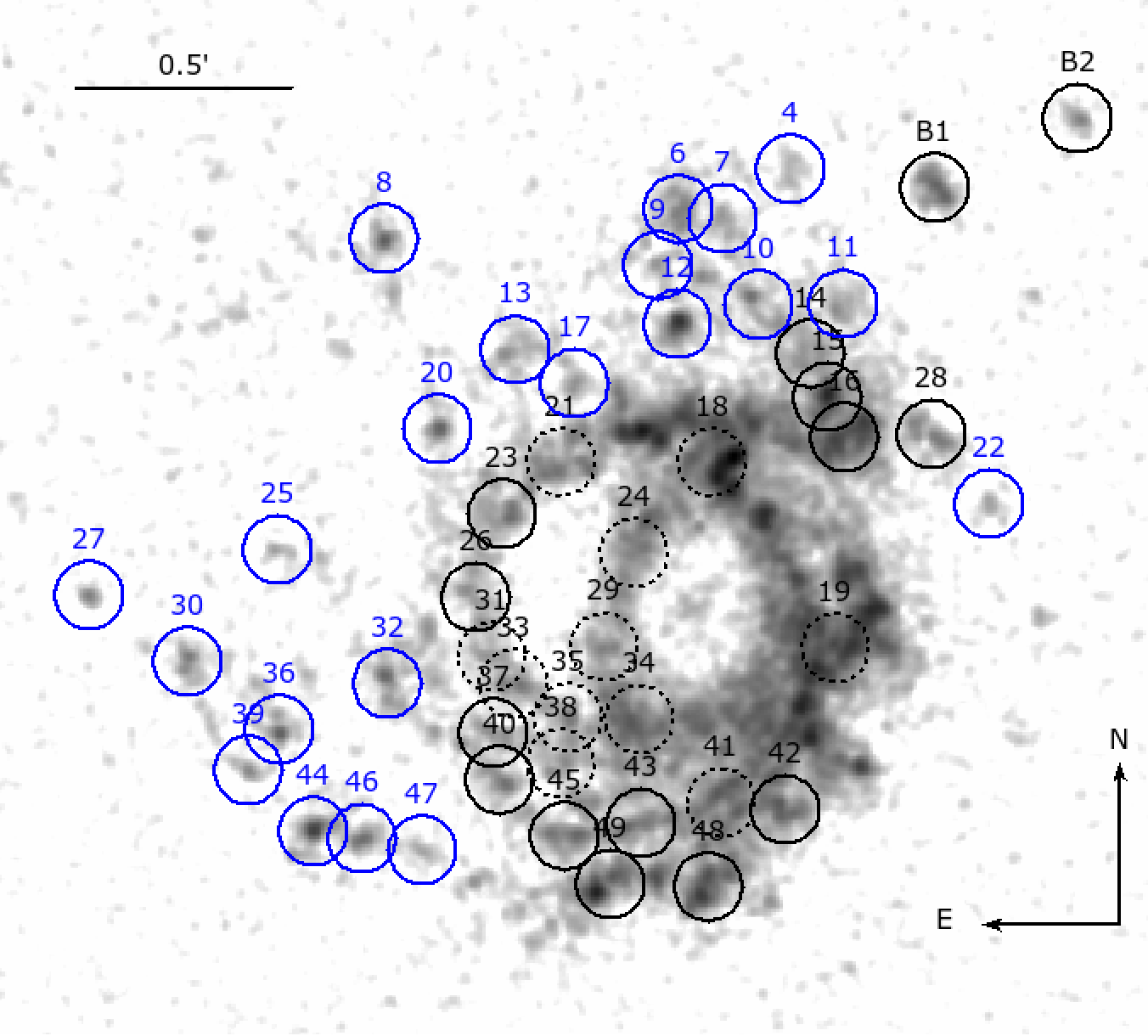}}}\hfil
  \subfigure[]{\includegraphics[scale=0.28]{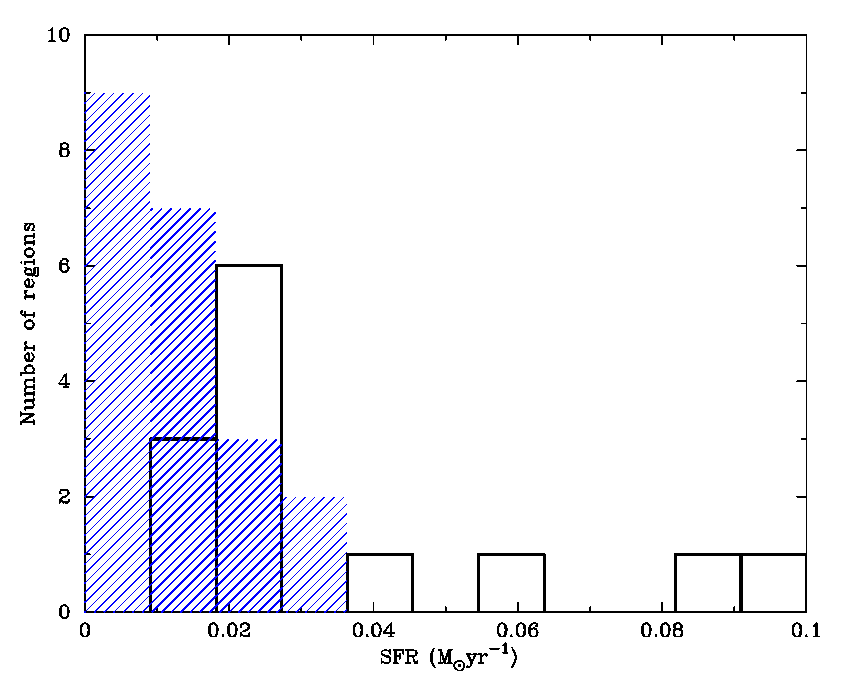}}
  \subfigure[]{\includegraphics[scale=0.28]{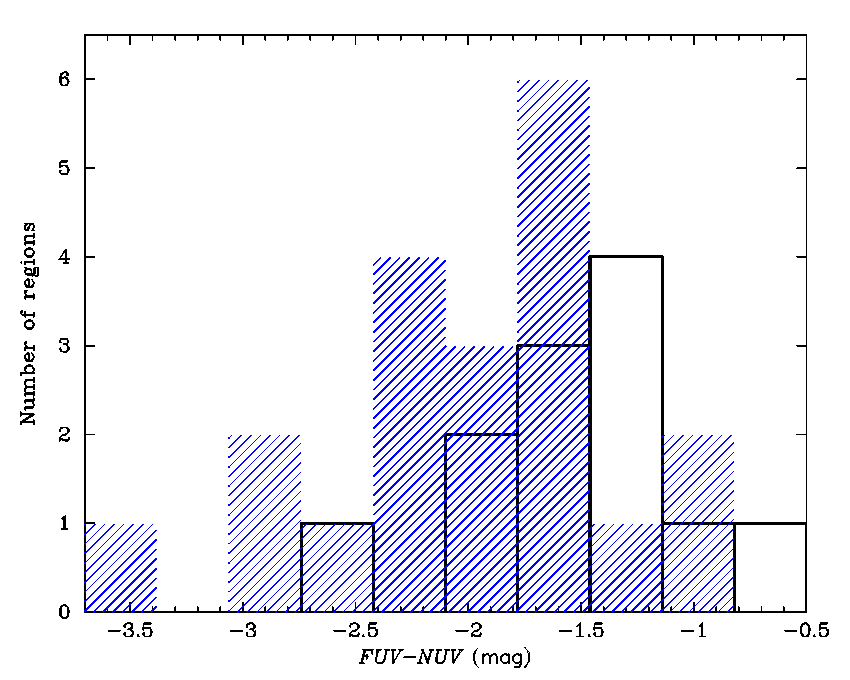}}\hfil
  \subfigure[]{\includegraphics[scale=0.28]{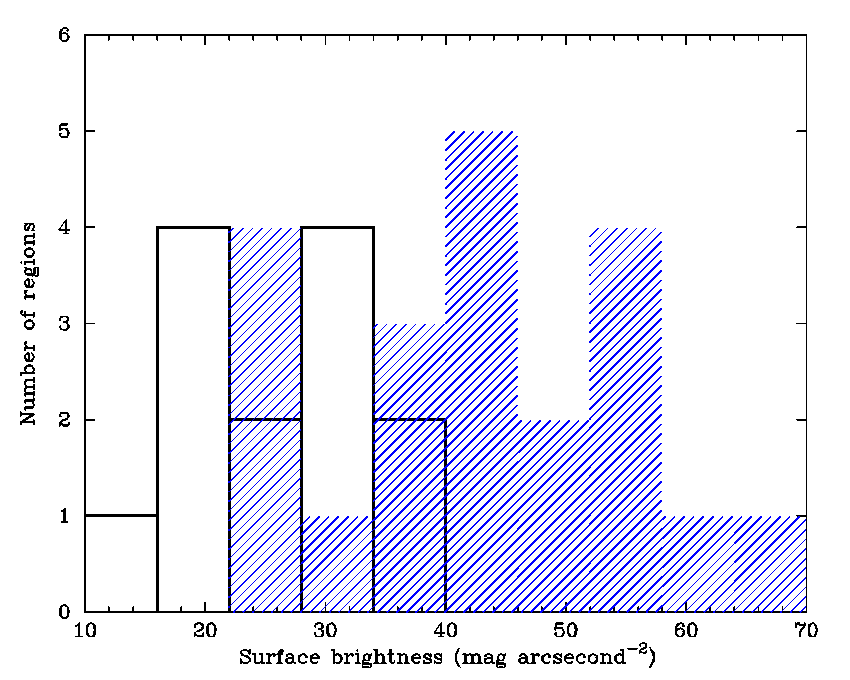}}\hfil

 \caption{(a) Individual `regions' detected by the {\sc sextractor} overlaid on the UVIT image are shown. The regions are colour-coded as follows:
 the regions with little to no optical emission are classified as XUV regions shown in {\it (blue)}, those which have optical emission in the $g$-band, and lie $> 0.5^\prime$ 
 away from the optical centre as regions in galaxy {\it (solid black)}, respectively. Objects marked as B1 and B2 are background galaxies and not considered in our analysis. 
 The counterparts of the latter form the bulk of the galaxy photometry in almost all the filters (and hence 
 are very different from the smaller star-forming regions/clumps explored in this analysis) are shown as {\it (dashed black)} circles. 
 The other panels show a comparison between the (b) SFR determined from {\it FUV} luminosity, (c) the {\it FUV-NUV} colour, and (d) surface brightness of the 
 regions in the galaxy {\it (solid black)}, and the XUV regions {\it (hatched blue)}, respectively. 
 It is evident that the XUV regions are bluer, brighter and have lower SFR relative to similar regions closer to the centre of the galaxy.}
 \label{regions}
 \end{figure*}

 It has been shown that the detection of extended UV (XUV) emission beyond the optical extent of the galaxy is not very uncommon in nearby star-forming galaxies \citep[e.g.][]{thilker07}. 
 A comparison between this star formation far away from the centre of the galaxy, and that in the optical disk is therefore crucial in understanding star formation processes. 
 Moreover, we are unaware of any XUV disk observed in clusters, which makes \ugc~a very unique galaxy. 
 
 We take advantage of the high resolution of the UVIT to get a better perspective on the individual regions which have been resolved in \ugc. In order to do so, we classified these regions 
 based on their distance from the centre of the galaxy. As shown in Fig.~\ref{regions}(a), all regions $>0.5^\prime$ from the centre of the galaxy are classified as XUV 
 regions (blue circles), and are mostly located outside the optical disk of the galaxy. Most of these XUV regions have a faint or no optical counterpart in the SDSS $g$-band image. 
 The regions which are part of the galactic disk are further sub-classified into two categories: regions which are comparable to the XUV regions in size and luminosity, and having an optical 
 counterpart are shown in solid black circles; these are henceforth referred to as the inner regions. Secondly, the regions closer to the centre of the galaxy, which form the bulk 
 of the emission from \ugc~in almost all filters are represented by dashed black circles. 
 In total there are 21 XUV regions which are compared to 13 inner regions, while the 11 innermost regions are ignored in the following analysis 
 because they are much more luminous (and massive) than the XUV regions. The objects marked as B1 and B2 are background galaxies, and are ignored hereafter.
 
 {\it FUV} light is emitted by stars with masses of several solar masses, and a mean age of 10 Myr ($10-100$ Myr), with the shortest wavelengths corresponding to the shortest
 timescales \citep{hao11,murphy11,kennicutt12}. The UV continuum emission from 1200 to 3200 \AA\ is therefore an inevitable consequence of recent star 
 formation. As a result, {\it FUV} luminosity is a direct tracer of recent star formation (few tens of Myr) occurring in a galaxy, and can be used to estimate the rate of star 
 formation (SFR).  In this work we employ the prescription provided by \citet{iglesias06}:
 \begin{equation}
 {\rm log(SFR}_{FUV}/{\rm M}_\odot ~yr^{-1}) = {\rm log}(L_{FUV}/L_\odot) - 9.51
 \end{equation}
 where, $L_{FUV}$ is the {\it FUV} luminosity of the galaxy corrected for galactic extinction ($A_{FUV}$). The calibration for this prescription is derived from Starburst99 models  
 \citep{leitherer99}, assuming a solar metallicity and a Salpeter IMF from $0.1-100~{\rm M}_\odot$. Galactic extinction, $A_{FUV} = 8.06 \times E(B-V)$ \citep{bianchi11}, 
 where the colour excess $E(B-V) = 0.0085$ for \ugc~is obtained from the dust reddening maps of \citet{schlafly11}\footnote{https://irsa.ipac.caltech.edu/applications/DUST/}.  

 Fig.~\ref{regions}(b) shows a comparison between the SFR estimated for the XUV regions, and the inner regions using the SFR derived from UVIT data. It is notable 
 that the XUV regions on average have lower SFR and smaller dispersion relative to similar regions closer to the centre of the galaxy. Specifically, the mean and standard 
 deviation of SFR for the XUV regions is $0.014\pm0.008~{\rm M}_\odot ~yr^{-1}$, relative to the value for the inner regions $0.035\pm0.028~{\rm M}_\odot ~yr^{-1}$. 
 The Kolmogorov-Smirnov (KS) statistical
 test probability ($p$), in favour of the hypothesis that the two distributions are drawn from the same parent sample is 1.77E-03, thus suggesting that they are disparate.    
 Furthermore, the XUV regions have bluer $FUV-NUV$ colours (Fig.~\ref{regions}(c); $p =$ 2.99E-02), and are brighter than the inner regions (Fig.~\ref{regions}(d); $p =$ 2.09E-03). 
 The respective mean surface brightness value for the XUV regions is $42.8\pm12.1$ magnitude arcseconds$^{-2}$, relative to only $26.1\pm7.2$ magnitude 
 arcseconds$^{-2}$ for the inner regions. 
 
  It is noteworthy that even though a correction for internal extinction is important when converting {\it FUV} luminosity to SFR \citep[e.g.][]{mahajan19}, in this case we 
 are unable to do so because of the unavailability of {\it FIR} data, and hence the estimated SFRs may be regarded as lower limits.  
 The result from these comparisons therefore should only be considered in a qualitative sense, assuming that the internal extinction in the XUV regions and inner regions is the same.
 Also, resolution effects may play a significant role in this comparison, since only the most massive star-forming regions may be resolved at the distance of \ugc. Most of the `regions' 
 considered here, therefore may be much larger than typical star-forming regions observed in nearby galaxies.  

 This distinction suggests that the XUV regions are statistically different from the inner regions. In our opinion, these qualitative differences 
 suggest that they have different origins. As \ugc~interacts with the surrounding ICM, some of the cold atomic gas from the galaxy is stripped, leaving a wake in the direction 
 opposite to the direction of the motion of the galaxy. The fact that the cluster centre lies diametrically opposite to the direction of the XUV tail(s) pointing towards the north-east, 
 forms the basis of this idea. Star-forming regions beyond the optical disk have been observed in several other nearby galaxies \citep[e.g.][]{cortese07,werk10,poggianti17,boselli21}. In 
 particular, the galaxy IC 3476 in the Virgo cluster which is experiencing edge-on stripping is worth mentioning, since it shows HII regions in the two diametrically opposite ends of the 
 galaxy just like \ugc~\citep[see fig. 4 of][]{boselli21}.
 Also, even though we only have access to the low-frequency radio data for \ugc, it is evident from Fig.~\ref{rad-g-uvit}, that there is more radio emission
 on the trailing edge of the galaxy relative to the western side facing the cluster centre. In this situation, the inner regions are formed where the stripped gas experiences 
 a shock, while the XUV regions are formed in the wake of the RPS gas tails. 


 
\begin{figure}
\begin{centering}
\includegraphics[scale=0.28]{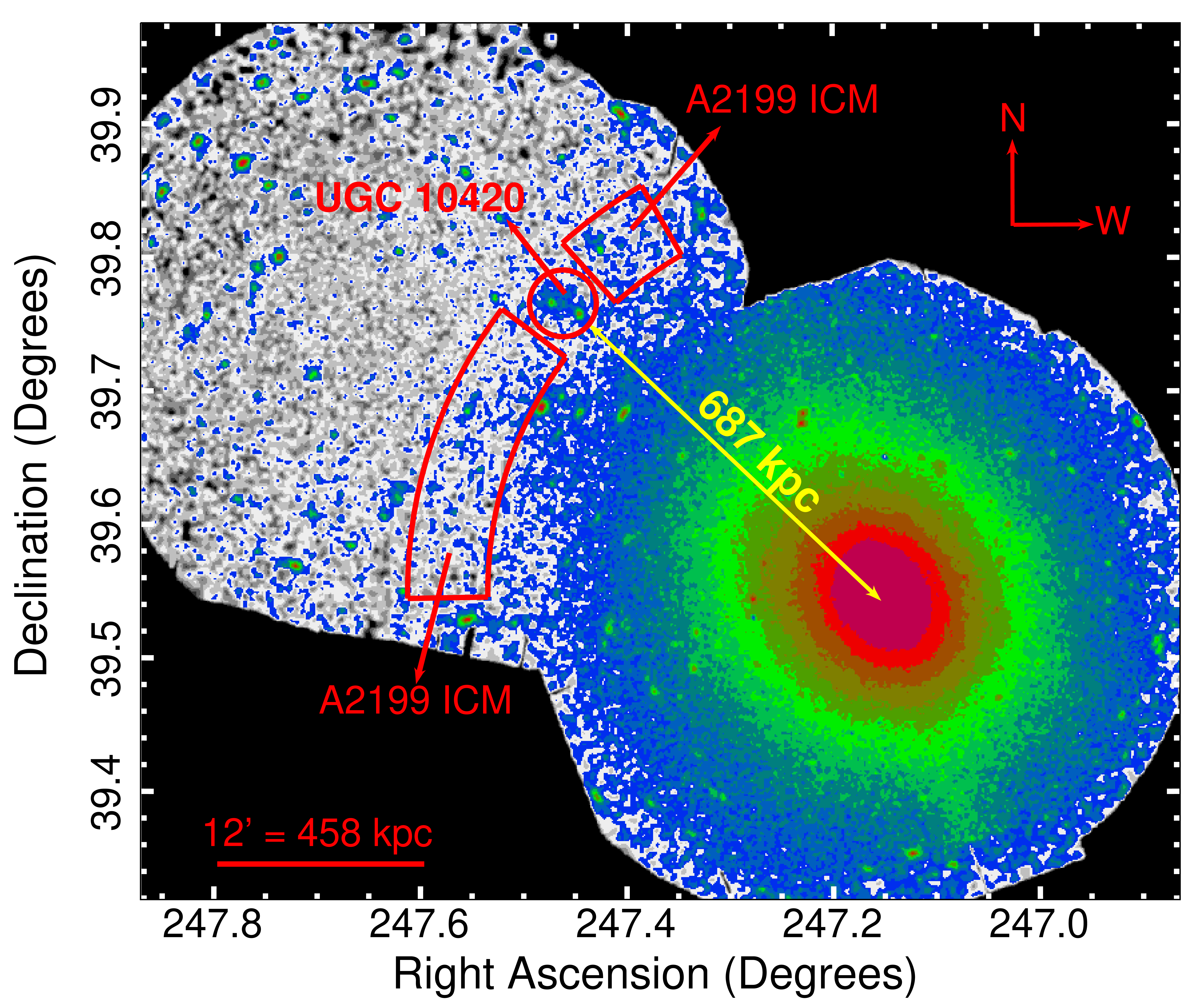}
\caption{X-ray image in the energy band 0.4--2.0 keV showing the location of UGC 10420 within its host galaxy cluster Abell 2199. The galaxy is located at a projected distance of 
 $\sim 680$ kpc from the X-ray peak of Abell 2199. The \textit{small red circle} (labelled \textit{UGC 10420}) is used for estimating the properties of the diffuse X-ray emission associated with UGC 10420. 
The partial annular regions sandwiching UGC 10420 (highlighted in \textit{red} colour and labelled \textit{A2199 ICM}) are used for estimating the properties of the Abell 2199 ICM at the 
location of the galaxy.}
\label{fig:a2199ugcloc}
\end{centering}
\end{figure}

 \subsection{Diffuse X-ray emission from UGC 10420}
 \label{xray}
 In \S\ref{xdata} we pointed out the existence of diffuse soft X-ray emission connecting two bright point-like sources and the nucleus of the galaxy (Fig.~\ref{fig:ugc10420xray} \textit{left panel}). Here, we have tried to isolate the diffuse component in UGC 10420 and study its properties. The properties of the point sources are studied and presented in the next section. In Figure 10, we show the X-ray emission from Abell 2199 and its outskirts.  The location of UGC 10420 with respect to Abell 2199 is also shown.  This figure was derived from both the current observation and another \textit{XMM-Newton} observation centered on the cluster centre. To isolate and study the properties of the diffuse X-ray emission associated with the galaxy, we have to estimate and remove the contribution of the backgrounds, cluster emission and the point sources.
 
We performed spectral analysis of emission within the red circular region (radius: $1.5\arcmin$; label: UGC 10420) shown in \autoref{fig:a2199ugcloc} to estimate the properties of overall diffuse X-ray emission from UGC~10420. The contribution from the two point sources in UGC 10420, shown in \autoref{fig:ugc10420xray}, was removed using the mask image generated by the task \textit{cheese} (Sec.~\ref{xdata}). The source spectrum, the modelled background spectrum, and the spectral response of the EPIC were extracted within the red circular region encompassing the diffuse emission in UGC 10420 using the tasks \textit{pn-spectra} (\textit{mos-spectra}) and \textit{pn$\_$back} (\textit{mos$\_$back}). All spectra were suitably grouped using the \texttt{FTOOL} \textit{grppha} (\texttt{HEASoft} version 6.30.1), and analysed using \texttt{XSPEC} version 12.12.1 \citep{xspec1996}. The diffuse background components towards UGC 10420 include the diffuse soft X-background (SXRB) and the cosmic X-ray background (CXB) which are described in detail in \citet{a2151}. 
 Since UGC 10420 is located in the outskirts of Abell 2199, there is also a significant contribution from the intracluster gas. Therefore, in addition to the X-ray background components, it is also necessary to account for the emission from the ICM of Abell 2199 at the galaxy's location, when modelling its diffuse emission, as described below. 


The partial annular regions sandwiching UGC~10420 shown in \autoref{fig:a2199ugcloc} (label: \textit{A2199 ICM}) were used to estimate the properties of the intracluster gas in the region around the galaxy. 
The PN, MOS1, and MOS2 spectra extracted from these annular regions were fitted jointly in the energy range 0.4--7.0 keV with the \textit{Rosat All Sky Survey} (RASS) diffuse background spectrum using the model \textit{constant*(apec+tbabs*(apec+apec+pegpwrlw) + tbabs*apec)}. We refer the reader to \citet{a2151} for definitions of the individual \texttt{XSPEC} models and details of the background modeling. For completeness, we mention that the model component \textit{apec+tbabs*(apec+apec+pegpwrlw)} in the expression above represents the complete X-ray background (SXRB + CXB). The second model component \textit{tbabs*apec} represents the absorbed X-ray emission from the cluster outskirts. The \textit{constant} factor is used to scale the model normalization of the PN, MOS1, and MOS2 data to ensure consistency with the RASS spectral data which is in units of cts s$^{-1}$ arcmin$^{-2}$. We fixed the total Hydrogen column density, $N_H$, along the line of sight to UGC 10420/Abell 2199 outskirts to $9.63\times 10^{19}$ cm$^{-2}$, based on the work of \citet{willingale2013}\footnote{\url{https://www.swift.ac.uk/analysis/nhtot/}}. The redshift was frozen at 0.03185. Two additional models -- \textit{gauss} (1.49 keV) and \textit{gauss} (1.75 keV) -- were used to model the instrumental background lines -- the Al K$\alpha$ line at $\sim$1.49 keV for PN, MOS1, and MOS2, and the Si K$\alpha$ line at $\sim$1.75 keV for the MOS1 and MOS2 detectors -- as and when required. 

The intracluster gas at the location of UGC 10420 was found to have a temperature of $3.3^{+0.9}_{-0.7}$ keV and metallicity of $\sim$0.2 Z$_{\odot}$ ($\chi^{2}=366.07$, $\text{dof}=354$). These properties of the ICM of Abell 2199 are in agreement with previous studies by \citet{a2199suzaku2010} and \citet{2020a2199}.\\

The properties of the diffuse X-ray emission associated with UGC 10420 were then estimated by fitting the PN, MOS1, and MOS2 spectra extracted from the red circular region in \autoref{fig:a2199ugcloc} simultaneously with the RASS background data using the model \textit{constant*(apec+tbabs*(apec+apec+pegpwrlw) + tbabs*(apec+apec))}. 
The newly added \textit{apec} component represents emission from the hot interstellar gas in the galaxy. 
The other model components have the same definitions as described previously. 
Here, the spectral parameters for the X-ray background, and the temperature and abundance of the Abell 2199 ICM were fixed based on the results of the spectral analysis mentioned above. The energy range 0.4--7.0 keV was used for the spectral fitting. We found the hot gas in UGC 10420 to have a temperature of $0.24^{+0.09}_{-0.06}$ keV, with the metal abundance, Z, frozen at the solar value ($\chi^2=104.53$, dof$=114$). If the abundance was left free to vary during the fit, we obtained Z $=0.8$ Z$_{\odot}$ with the gas temperature unchanged, but the errors in the abundance could not be constrained due to low number of counts.

 \subsection{X-ray emission from the two hot spots and nuclear region}
 \label{sec:xrayspecptsrc}

\begin{figure}
\begin{centering}
\includegraphics[scale=0.28]{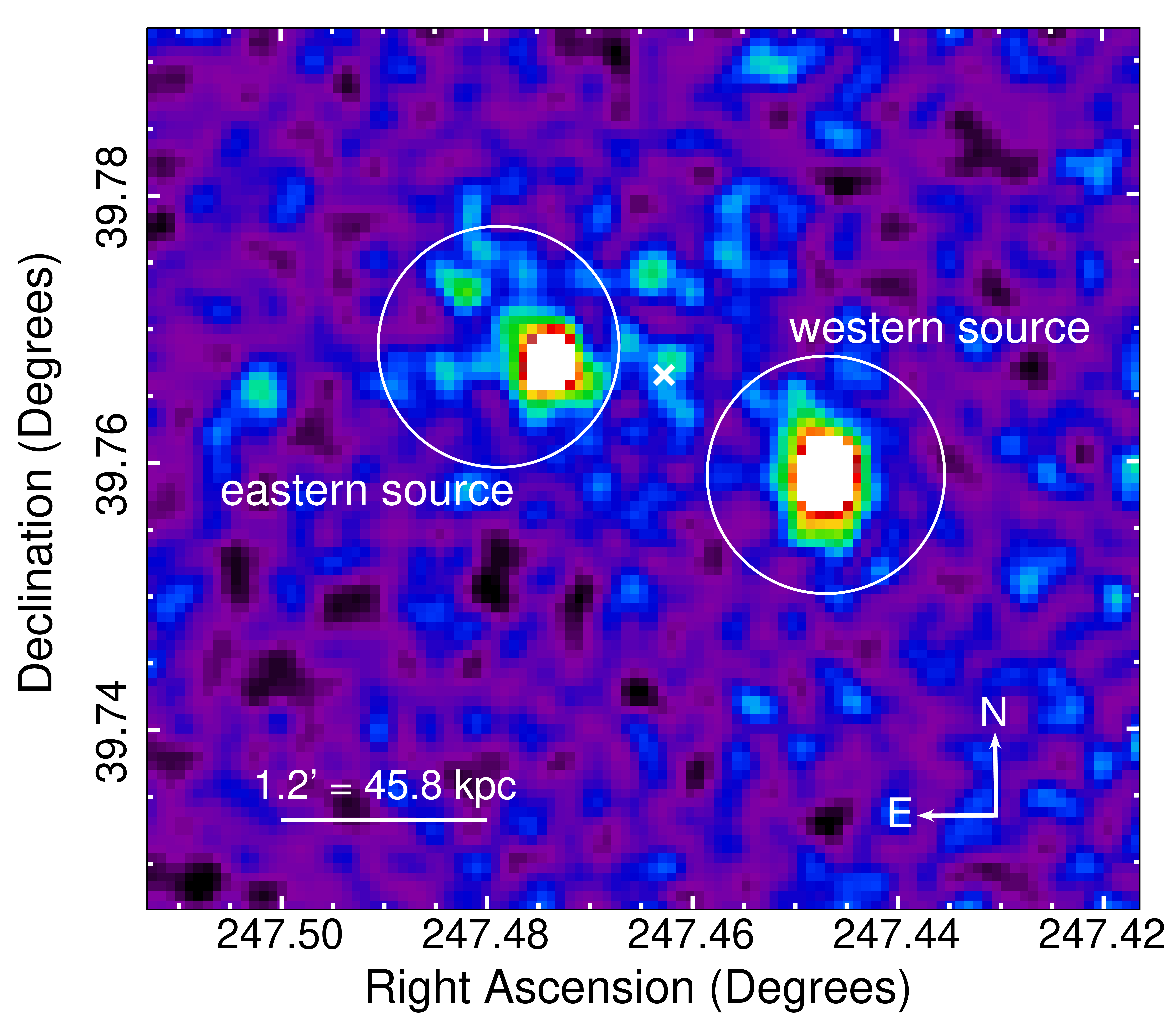}
\caption{The total band (0.4--7.0 keV) X-ray image showing the eastern and western X-ray point sources in \ugc. The ${\bf \times}$ symbol marks the optical centre of UGC 10420. The \textit{solid white circles} mark the regions used for the spectral extraction and analysis of each point source.}
\label{fig:ptsrcspecimg}
\end{centering}
\end{figure}

Spectral analysis of the two bright point sources in UGC 10420 was carried out by extracting spectra from the circular regions shown in \autoref{fig:ptsrcspecimg} to estimate their general properties. 
These relatively large regions, extending beyond the actual size of the detected sources, were used so as to enclose enough counts to make the spectral modelling possible. 
Since these circular regions also enclose the diffuse emission from the cluster and galaxy, we accounted for these contributions in the spectral analysis as described above.
The spectrum of each point source was well fitted by a power-law\footnote{The power-law photon spectrum has the form $KE^{-\alpha}$, where $\alpha$ (dimensionless) is the photon index of the power-law, and $K$ is the normalization in units of photons keV$^{-1}$ cm$^{-2}$ s$^{-1}$ at 1 keV.} in the energy range 0.4--7.0 keV. 
The unabsorbed flux (luminosity) of each point source was estimated by convolving the \textit{powerlaw} model with the \texttt{XSPEC} model \textit{cflux} (\textit{clumin}) after freezing the power-law model normalization. 
The results of the spectral analysis and the derived properties of the two point sources are given in \autoref{tab:ptsrctable}.  
 

\begin{table*}\setlength{\tabcolsep}{6pt}
\begin{center}
\caption{Properties of the two X-ray point sources in UGC 10420. The errors in photon index, flux, and luminosity are quoted at 90 per cent confidence level.}
\begin{tabular}{ccccccc}
\hline
\hline
&R.A. (J2000)& Dec. (J2000)& Photon index ($\alpha$) &$\chi^{2}$ (dof)& Flux & Luminosity\\
&&&&&(0.4--7.0 keV)&(0.4--7.0 keV)\\
&&&&&erg cm$^{-2}$ s$^{-1}$&erg s$^{-1}$\\
\hline
Eastern source &16h29m53.92s&+39d46m01.67s&$1.8^{+0.3}_{-0.2}$&20.45 (34)& $2.6\pm{0.4} \times 10^{-14}$& $6.7\pm{1.1} \times 10^{40}$\\
Western source &16h29m47.41s&+39d45m31.78s&$1.6\pm{0.2}$&61.13 (58)& $4.9\pm{0.5} \times 10^{-14}$&$1.2\pm{0.1} \times 10^{41}$\\
\hline
\hline
\end{tabular}
\label{tab:ptsrctable}
\end{center}
\end{table*}

The unabsorbed X-ray flux and luminosity of the entire galaxy (hot spots + diffuse emission) were estimated within two circular apertures of radii $0.8\arcmin$ and $1.5\arcmin$ centred at the optical centre of the galaxy (see Figure \ref{fig:ugc10420xray}). In order to do so, the PN, MOS1, and MOS2 spectra extracted within the $1.5\arcmin$ radius aperture were simultaneously fitted with the RASS data using the model \textit{constant*(apec+tbabs*(apec+apec+pegpwrlw) + tbabs*(apec+  {\bf apec + powerlaw + powerlaw}))}. 
 The highlighted model component 
 represents the total X-ray emission -- point sources + diffuse -- from UGC 10420. Here, \textit {apec} represents the diffuse emission from UGC 10420, and the two \textit {powerlaw} components model the emission from the two point sources in UGC 10420 which lie within the aperture of radius $1.5\arcmin$. For modelling the X-ray spectrum within the smaller aperture of radius $0.8\arcmin$, we used only one \textit{powerlaw} model since only the eastern point source falls within this aperture. The X-ray flux in the two apertures was then estimated by convolving the \textit{apec + powerlaw + powerlaw} component with the \texttt{XSPEC} model \textit{cflux}. Similarly, the total X-ray luminosity was estimated using the \texttt{XSPEC} convolution model \textit{clumin}. The total X-ray flux and luminosity values for the two apertures in the soft, hard, and total bands are reported in \autoref{tab:xrayflux}. 

 A limit to the soft X-ray flux from the nuclear region, which is not fully resolved, was estimated from the PN data in the soft band and using a circular region with a radius of 9 arcsec. Scaling the estimated PN background to the same size, we obtain a 3$\sigma$ upper limit on the soft emission from the nuclear region around the optical centre  to be $\sim$0.004 counts s$^{-1}$. Assuming the central source to have a photon index of 2, this limit translates to a flux of 5.4$10^{-15}$ erg cm$^{-2}$ s$^{-1}$ in the soft energy band, giving a 3$\sigma$ limit on luminosity of 1.4 $10^{40}$ ergs s$^{-1}$ for an AGN like source in the centre.


 \begin{table*}\setlength{\tabcolsep}{6pt}
\begin{center}
\caption{Total unabsorbed X-ray flux ($F_X$ in erg cm$^{-2}$ s$^{-1}$) and luminosity ($L_X$ in erg s$^{-1}$) estimates of UGC 10420. The errors are quoted at 90 per cent confidence level. }
\begin{tabular}{cccc}
\hline
\hline
Aperture radius&Soft band & Hard band & Total band\\
&(0.4--2.0 keV)&(2.0--7.0 keV)&(0.4--7.0 keV)\\
\hline\\
$1.5\arcmin$&$F_X$: $3.7\pm{0.5}\times 10^{-14}$&$F_X$: $3.6^{+0.4}_{-0.5}\times 10^{-14}$&$F_X$: $7.3\pm{0.9}\times 10^{-14}$ \\
&$L_X$: $9.3\pm{1.1}\times 10^{40}$ &$L_X$: $9.1\pm{0.9}\times 10^{40}$ &$L_X$: $1.8\pm{0.2}\times 10^{41}$ \\
\\
$0.8\arcmin$&$F_X$: $2.5\pm{0.3}\times 10^{-14}$ &$F_X$: $2.2\pm{0.3}\times 10^{-14}$ &$F_X$: $4.7\pm{0.6}\times 10^{-14}$\\
&$L_X$: $6.2\pm{0.7} \times 10^{40}$ &$L_X$: $5.7^{+0.6}_{-0.8}\times 10^{40}$ &$L_X$: $1.2\pm{0.1}\times 10^{41}$ \\
\hline
\hline
\end{tabular}
\label{tab:xrayflux}
\end{center}
\end{table*}

 \section{Discussion}
 \label{results}
 
 In this paper we have presented the UVIT {\it FUV} data and ancillary data in other wavebands for the spiral galaxy \ugc, a member of the cluster Abell 2199. Our analysis shows that this galaxy, 
 at a projected distance of $\sim 680$ kpc from the cluster's centre is being ram-pressure stripped by the ICM of the cluster. It is notable that the l.o.s. velocity offset of \ugc~with respect to the mean 
 velocity of Abell 2199 is only $\sim 372$ km s$^{-1}$, which is much smaller than the velocity dispersion of the cluster. This, together with the \fuv observations presented in this work, suggests that  
 \ugc~must be moving with a very high speed towards the cluster centre, in a direction normal to the l.o.s.. As a consequence of stripping, the outer spiral arms of the galaxy are 
 being removed and turned into what looks like a wake of star-forming knots on the south-eastern side of the galaxy (Fig.~\ref{uvit}). These observations agree with the simulations 
 of RPS galaxies suggesting that the SFR increases in the gaseous wake of the galaxy rather than the disk itself, and hence is strongly dependant on the density of the surrounding 
 ICM \citep{kapferer09}.  
  
 There is an increasing consensus building up in the recent literature that most of the enhanced star formation in RPS galaxies is observed on the `leading edge' of the galaxies
 \citep{gavazzi15,vulcani18,hess22,roberts22a}. 
 Such galaxies, showing an unambiguous tail of material stripped from the disk are often called the `jellyfish galaxies', and are considered to be a product of RPS in clusters. 
 It is believed that the stripped gas must cross the entire galaxy disk before leaving it, and therefore is more likely to cause
 instabilities and turbulence, which in turn triggers starburst in jellyfish galaxies \citep{boselli22}. Several examples of such jellyfish galaxies have been observationally discovered
 \citep{poggianti16,vulcani18,safarzadeh19,lee20,cramer21,roberts22}. The {\it FUV} observations of \ugc, showing intense star-forming knots on the western side of the disk 
 (Fig.~\ref{uvit}), also favours this scenario.  
 
  In order to test the significance of the increase in the SFR of \ugc, we adopted the following SFR main sequence relation derived by \citet{speagle14} as a function of stellar mass and 
 cosmic time: \\
  ${\rm log~SFR}(M^*,t) = (0.84-0.026t){\rm log}M^*-(6.51-0.11t)$, \\ 
 where, $t$ is the age of the Universe (Gyr) at the redshift of the galaxy. In line with many other observations in the literature \citep[][and references therein]{lee22}, we find that the 
 SED derived SFR of \ugc~is at least a factor of nine larger than that expected for star-forming field galaxies of similar mass at $z\sim0.03$.   
   
 The observations presented in this work are also in good agreement with a recent study of cluster galaxies in the EAGLE simulations, where \citet{eagle20} have shown that properties 
 such as the SFR, star formation efficiency and ISM pressure are enhanced on the leading edge of the normal star-forming galaxies. On the other hand, an increase in the gas particles is observed on 
 the `trailing edge' of these galaxies, suggesting removal of cold gas via RPS. Similar results were obtained by  \citet{boselli21}, who presented hydrodynamical simulations of 
 the RPS galaxy IC 3476 along with its multi-wavelength data. The fact that \ugc~seems to have more gas present on the trailing edge (Fig.~\ref{rad-g-uvit}), and an asymmetric 
 distribution of star-forming regions, supports the results from the simulations.
  
 These results from simulations were recently tested observationally by \citet{roberts22}, who 
 found them to agree with the resolved analysis of the integral field spectroscopic data of 29 jellyfish galaxies. A case study of IC 3949 presented by these authors is further evidence in favour of 
 a scenario where the increased ISM pressure on the leading edge compresses the atomic Hydrogen and turns it into molecular gas, thereby triggering a localized burst of star formation. 
 \citet{eagle20} also showed that this effect is most pronounced in galaxies in the stellar mass range of $10^{9.5}-10^{10.5}~M_{\odot}$, which conveniently includes 
 \ugc~(Table~\ref{magphys}).
 
 An examination of the individual XUV star-forming regions (defined in Sec.~\ref{s:regions}) shows that they are statistically brighter than their counterparts at lower galactocentric distances, 
 and have relatively bluer {\it FUV-NUV} colours and lower SFR. 
 These observations are in agreement with the results of \citet{werk10}, who analysed the HI and {\it FUV} data for some nearby star-forming galaxies. \citet{werk10} found that 
 8\%--11\% of emission-line point sources, near star-forming galaxies in their sample, are HII regions associated with the extended UV disk of the galaxies.  
 
 Any environmental mechanism which influences the gas content of a galaxy, thereby affecting its SFR, can potentially impact the AGN activity as well. In order to test this hypothesis, 
 \citet{peluso22} have recently exploited the integral field data for 115 RPS galaxies from MANGA and MUSE. \citet{peluso22} find that AGN fraction among RPS galaxies is higher than their 
 non-RPS counterparts. Furthermore, the incidence of optical AGN (classifiable using the BPT diagnostics) is $51\%$ among massive ($M^* \geq 10^{10} M_{\odot}$) RPS galaxies. 
 It therefore comes as no surprise that \ugc~is optically classified as a LINER \citep{toba14}. 
 On the contrary, evidence has been presented in the literature to show that presence of a low-ionization emission region may not always correspond to nuclear activity \citep[e.g.][]{belfiore16}. 
 Interested readers are directed to \citet[][and references therin]{boselli22}, for a detailed discussion on the subject, while deeper and high resolution data would be required to confirm 
 presence of AGN in \ugc.


 Analysis of archival X-ray data shows that UGC~10420 lies just at the edge of the X-ray emitting region of Abell~2199. Two bright X-ray sources, a few tens of kpc away from the centre of the galaxy without any optical counterpart are clearly visible, along with very faint emission from the central region of the galaxy.  Such bright sources (hot spots, hereafter),  
 are reminiscent of those generally observed around bright active galactic nuclei like Pictor A. Although such hot spots are often connected with a jet seen as faint diffuse 
 emission \citep{hardcastle}, such is not the case with \ugc. 
 The eastern hot spot appears to be embedded in diffuse radio emission at that location as has been seen in Pictor A \citep{thimmappa}. It should be noted that the nucleus in Pictor A is about 20 times brighter than the hot spots \citep{hardcastle}, while almost the reverse is true here. 
 
 The X-ray spectrum of the two hot spots in UGC 10420 are well fitted by a power-law with photon index of $1.8^{+0.3}_{-0.2}$ and $1.6\pm{0.2}$ for the eastern and western source, respectively. The luminosities of these hot spots are similar to the hot spots in Pictor A, while the nucleus is an order of magnitude weaker. If confirmed with deeper {\it Chandra} observations with higher resolution, this would be the first case of hot spots associated with such a weak nucleus. This also indicates that we do not find any evidence of enhancement in the nuclear activity in the galaxy due to the RPS process, which has a luminosity typical of low-activity nucleus in LINERS.  
 
 A careful analysis of diffuse X-ray emission indicates that the hot galactic plasma in \ugc~has a temperature of $0.24^{+0.09}_{-0.06}$ keV, consistent with that observed in spiral galaxies 
 \citep[e.g.][]{2008ngc5775,mineo2012}. Also, the X-ray luminosity of the diffuse gas component in \ugc~is $1.8\pm{0.9}\times 10^{40}$ erg s$^{-1}$ (0.4--2.0 keV), which is typical of spiral galaxies \citep{tyler2004, owen2009, mineo2012}. This also indicates that the RPS has no effect on the hot gas in the galaxy.

   
 In a nutshell, multi-wavelength analysis of \ugc~confirms that it is a RPS galaxy which is undergoing a short-lived phase of intense star formation, as it interacts with the ICM in the cluster 
 Abell~2199. While the star-forming knots on the western side of the disk are formed as a consequence of the compression of ISM of the galaxy due to RPS, the star formation 
 on the eastern side is part of the wake which comprises loosely bound or ejected gas from the galaxy. References from the literature suggest that deep narrow-band and radio continuum 
 imaging of \ugc~should produce further evidence to support the ideas presented in this paper. Although it is worth mentioning that \ugc~is unique not only because of the asymmetric 
 distribution of star-forming regions, but also due to its status as an XUV disk galaxy hosted by a cluster. 
   
 \section{Epilogue}
 \label{epilogue}
 
 Although the multi-wavelength analysis of the data for \ugc~has thrown some light on the evolution of this and such similar objects, it is worth mentioning the paucity of 
 narrow-band data. We believe narrow-band imaging (H$\alpha$), sensitive to the presence of massive O stars would be a key to probing the XUV star-formation in this 
 galaxy. Also, continuum and HI emission, and molecular gas observations can throw light on the distribution of gas in and around the galaxy, and may prove very helpful in understanding 
 the role of RPS in the evolution of \ugc, just like other RPS galaxies observed in clusters \citep[e.g.][]{smith10,werk10,cortese07,yagi10,roman19,boselli21}. 
 
\section{Acknowledgements}
 
 S. Mahajan was funded by the SERB Research Scientist (SRS) award (SB-SRS/2020-21/56/PS), Department of Science and Technology (DST), Government of India.  
 Kulinder Pal Singh thanks the Indian National Science Academy for support under the INSA Senior Scientist Programme. Authors are grateful to Dr. Dharam Vir Lal for his help 
 with  the {\it LOFAR} data. We are grateful to the anonymous reviewer and the associate editor for their constructive criticism which helped in improving the readability of this manuscript. 
  
 This publication uses data from the {\it AstroSat} mission of the Indian Space Research Organisation (ISRO), archived at the Indian Space
 Science Data Centre (ISSDC). UVIT project is a result of collaboration between IIA (Bengaluru), IUCAA (Pune), TIFR (Mumbai), several  
 centres of ISRO, and the Canadian Space Agency (CSA). 
 This research has made use of the NASA/IPAC Infrared Science Archive, which is funded by the National Aeronautics and Space Administration and 
 operated by the California Institute of Technology.  The {\sc topcat} software \citep{taylor05} was used for some of the analysis presented in this paper.  
 This publication makes use of data products from the Wide-field Infrared Survey Explorer, which is a joint project of the University of California, Los Angeles, and the 
 Jet Propulsion Laboratory/California Institute of Technology, funded by the National Aeronautics and Space Administration.
 This dataset or service is made available by the Infrared Science Archive (IRSA) at IPAC, which is operated by the California Institute of Technology under contract with the 
 National Aeronautics and Space Administration.  
 We acknowledge the use of LOFAR data in accordance with the credits given on \url{https://lofar-surveys.org/credits.html}, and SDSS data given on 
 \url{https://www.sdss.org/collaboration/citing-sdss/}. The X-ray data were downloaded from the High Energy Astrophysics Science Archive Research Center (HEASARC), 
 maintained by NASA’s Goddard Space Flight Center. 
 This research has made use of SAOImageDS9, developed by Smithsonian Astrophysical Observatory, and the HEASoft FTOOLS (http://heasarc.gsfc.nasa.gov/ftools).

\bibliographystyle{pasa-mnras}
\bibliography{u10420-v7}

\end{document}